\documentclass[11pt]{article}

\usepackage{graphicx}
\usepackage{caption}
\usepackage{subcaption}
\usepackage{lineno}
\usepackage{float}

\usepackage{verbatim}
\usepackage{amsmath}
\usepackage{amsfonts}
\usepackage{amssymb}
\usepackage{color}
\usepackage{hyperref}
 \usepackage{stackrel}
\usepackage{upgreek}
 \usepackage{csquotes}
 \usepackage{ulem}
 \usepackage{cite}

 \usepackage{fullpage}

 \def\##1{{\bf #1}}
\renewcommand\vec{\mathbf}

\def\=#1{\underline{\underline{#1}}}

 \def\eps{\varepsilon}

 \def\epso{\eps_{\scriptscriptstyle 0}}

\def\epsr{\eps_{\rm r}}

\def\ux{\hat{\#x}}
\def\uy{\hat{\#y}}
\def\uz{\hat{\#z}}
\def\ur{\hat{\#r}}
\def\up{\hat{\#p}}

\def\fsca{f_\text{pert}}
\def\Fsca{F_\text{pert}}

\def\bro{{\#r}_{\rm o}}
\def\ro{r_{\rm o}}
\def\thetao{\theta_{\rm o}}
\def\phio{\phi_{\rm o}}

\def\.{\mbox{ \tiny{$^\bullet$} }}

\def\les{\left[}
\def\ris{\right]}
\def\lec{\left\{}
\def\ric{\right\}}

\def\calA{{\cal A}}
\def\calB{{\cal B}}
\def\calC{{\cal C}}
\def\calD{{\cal D}}
\def\calI{{\cal I}}
\def\calJ{{\cal J}}
\def\calK{{\cal K}}
\def\calL{{\cal L}}
\def\calM{{\cal M}}
\def\calN{{\cal N}}
\def\calT{{\cal T}}

\def\alphax{\alpha_{\rm x}}
\def\alphay{\alpha_{\rm y}}

\def\balpha{\bar{\alpha}}
\def\rtilde{\tilde{r}}

\def\one{^{(a)}}
\def\two{^{(b)}}

\def\inc{_{\rm source}}
\def\sca{_{\rm pert}}
\def\ind{_{\rm int}}

\def\smn{_{\rm smn}}
\def\smnp{_{\rm s^\prime m^\prime n^\prime}}
\def\double{_{\rm ss^\prime mm^\prime nn^\prime}}

\def\emn{_{\rm emn}}
\def\omn{_{\rm omn}}

\def\Emn{E_{\rm mn}}

\begin{document}

\begin{center}
\textbf{Theory of perturbation of electrostatic field by an  {anisotropic}  dielectric sphere} \\[10pt]
\textit{Akhlesh Lakhtakia}\\
{Department of Engineering Science and Mechanics, The Pennsylvania State University, University Park, Pennsylvania 16802, USA}\\
 \textit{Nikolaos L. Tsitsas}\footnote{Corresponding author; e-mail: ntsitsas@csd.auth.gr}\\
{School of Informatics, Aristotle University of Thessaloniki, Thessaloniki 54124, Greece}\\
\textit{Hamad M. Alkhoori}\\
{Department of Electrical Engineering, United Arab Emirates University, P.O. Box 15551, Al Ain, UAE}

\end{center}

\begin{abstract}

The boundary-value problem for the perturbation of an electric potential by a homogeneous  {anisotropic}  dielectric sphere in vacuum
was formulated. The total   potential in the exterior region was expanded in series of radial polynomials and tesseral harmonics, as is standard for the Laplace equation. A bijective transformation of space was carried out to formulate a series representation of the potential in the interior region. Boundary conditions on the spherical surface were enforced to  derive a transition matrix that relates
the expansion coefficients of the perturbation potential in the exterior region to those of the source potential. Far from the sphere, the perturbation potential decays as the inverse of the distance squared from the center of the sphere, as confirmed numerically.
\end{abstract}

\section{Introduction}

 When an object made of a certain linear homogeneous dielectric material is exposed to a time-invariant electric field, the atoms constituting the object interact with that electric field until an electrostatic steady state is reached a short time later  \cite{Smythe,Jackson}. This interaction is described in terms of the polarization $\#P$ which is   the volumetric density of    electric dipoles induced inside the object. The polarization $\#P$ is linearly proportional of the electric field $\#E$, the proportionality constant being the dielectric susceptibility of the material multiplied by the free-space permittivity. Coulomb's law then leads to the definition of the electric displacement field $\#D$ which is linearly related to $\#E$ by the permittivity of the material. The permittivity is scalar for isotropic materials but dyadic for anisotropic materials \cite{Gersten,Nye}.

 Perturbation of an electrostatic field by a linear homogeneous dielectric object in free space has been studied for a long time \cite{Kellogg,Williams,Love,Jones,Lindell,Matt}, presently with biological \cite{Sukhorukov}, biomedical \cite{Woeppel},   electrochemical \cite{Everts}, and manufacturing \cite{Plog} applications.
  Let us refer to the electric field present in the absence of the object as the \textit{source field},
the difference between the electric field present at any location outside the object and the source electric field as the
the \textit{perturbation field}, and the electric field present at any location inside the object as the \textit{internal field}.
The boundary-value problem for the electrostatic steady state is solved analytically by (i) expanding the source, perturbation, and internal    fields in terms of suitable basis functions and (ii) imposing appropriate boundary conditions at the surface of the perturbing object. The basis functions suitable for representing the source and the perturbation fields are the eigenfunctions of the  Laplace equation \cite{Kellogg,Moon,Medkova}.
When the object is made of an isotropic material, the basis functions suitable for representing the internal field   also are the eigenfunctions
of the Laplace equation.

 {Our objective here is to show that analytic
 basis functions for representing
the internal field are available}
 when the perturbing
 object is composed of an  {anisotropic}  dielectric material \cite{Auld} described by
 the constitutive relation
\begin{equation}  \label{constitutive}
\#D(\#r)= \=\eps \. \#E(\#r)\,,
\end{equation}
where the permittivity dyadic
\begin{equation} \label{eps-def}
\=\eps = \epso\epsr\,\=A\.\=A\
\end{equation}
involves {
the diagonal dyadic
\begin{equation}
\label{A-def}
\={A}= \alphax^{-1}\,\ux \ux + \alphay^{-1} \,\uy \uy + \uz \uz\,
\end{equation}
with
$\epso$ denoting the permittivity of free space. Any real symmetric dyadic
can be written in the form of Eqs.~(\ref{eps-def}) and (\ref{A-def}) by virtue of the principal axis theorem \cite{Charnow,Strang}.
The only conditions imposed by us are that the scalars
$\alphax>0$, $\alphay>0$, and  $\epsr>0$.  Natural materials of such kind exist \cite{Auld}.
Materials of this kind  can also be realized as   homogenized composite materials}  by properly dispersing dielectric fibers  in some host isotropic dielectric material \cite{MAEH}.

The positive definiteness \cite{Lutkepohl} of $\={A}$  allows for an affine transformation of space in which the governing equation from which the eigenfunctions are obtained is transformed into the Laplace equation.
After solving the Laplace equation and obtaining the eigenfunctions in the transformed space,   an inverse transformation of space is effected
to  obtain  eigenfunctions in the original space for the material described by Eq. (\ref{constitutive}).  We apply the procedure here to analytically investigate the
perturbation of an electrostatic field by the  {anisotropic}  dielectric sphere.

The plan of this paper is as follows. Section~\ref{bvp} contains the formulation of the boundary-value problem in terms of
the electric potential. The expansion of the potential in the exterior region, which is vacuous, is presented in Sec.~\ref{bvp-ext}; two
illustrative examples of the source potential are provided in Sec.~\ref{bvp-sp};
the expansion of the potential inside the  {anisotropic}  dielectric sphere is derived
in Sec.~\ref{bvp-int}; boundary conditions are enforced in Secs.~\ref{bvp-bc}--\ref{bvp-rce} to derive a transition matrix that relates
the expansion coefficients of the perturbation potential to those of the source potential; the symmetries of the transition
matrix are presented in Sec.~\ref{bvp-symm}; and an asymptotic expression for the perturbation potential is derived in Sec.~\ref{bvp-asymp}. Section~\ref{nrd}   presents illustrative numerical results.

\section{Boundary-Value Problem}\label{bvp}
The region $r>a>0$ is taken to be vacuous with constitutive relation $\#D(\#r)=\epso\#E(\#r)$, whereas the spherical region $r<a$ is occupied by the chosen material described by Eq. (\ref{constitutive}).  As electrostatic fields always satisfy the relation $\nabla \times \vec{E}(\vec{r}) = \vec{0}$ \cite{Jackson}, it follows that $\vec{E}(\vec{r}) = - \nabla \Phi(\vec{r})$, where $\Phi(\vec{r})$ is the electric potential.
Henceforth, we use the electric potential.

\subsection{Potential in the region $r>a$}\label{bvp-ext}
The solution of the Laplace equation $\nabla^2\Phi(\#r)=0$ in the spherical coordinate system $\#r\equiv(r,\theta,\phi)$
has been known for almost two centuries;
thus \cite{Jackson,Moon},
\begin{equation}
\label{eq4}
\Phi(\#r)= \sum_{s\in\left\{e,o\right\}}^{}\sum_{n=0}^{\infty}\sum_{m=0}^{n}\lec \Emn\les
\calA\smn\, r^n + \calB\smn \,r^{-(n+1)}\ris\, Y\smn(\theta,\phi)\ric\,,\quad r > a\,,
\end{equation}
where
\begin{equation}
\Emn =  \left(2-\delta_{m0}\right)\frac{2n+1}{4\pi}\,\frac{(n-m)!}{(n+m)!}\,
\end{equation}
is a normalization factor with $\delta_{mm^\prime}$ as the Kronecker delta
and
the tesseral harmonics
\begin{equation}
\left.\begin{array}{l}
Y\emn(\theta,\phi) = P_n^m(\cos\theta) \cos(m\phi)
\\[5pt]
Y\omn(\theta,\phi) = P_n^m(\cos\theta) \sin(m\phi)
\end{array}\right\}\,
\end{equation}
involve the associated Legendre function $ P_n^m(\cos\theta)$  \cite{Morse,Alkhoori1}.   The coefficients $\calA\smn$ are associated
with terms that are regular at the origin, whereas the coefficients $\calB\smn$ are associated with terms
that are regular at infinity. The definitions of the tesseral harmonics mandate that $\calA_{\rm o0n}= 0$
and $\calB_{\rm o0n}= 0$~$\forall\,n$.

On the right side of Eq.~(\ref{eq4}),
\begin{equation}
\label{Phi-inc}
\Phi\inc(\#r)= \sum_{s\in\left\{e,o\right\}}^{}\sum_{n=0}^{\infty}\sum_{m=0}^{n}\les \Emn
\calA\smn \,r^n  \, Y\smn(\theta,\phi)\ris
\end{equation}
is the source potential whereas
\begin{equation}
\label{Phi-sca}
\Phi\sca(\#r)= \sum_{s\in\left\{e,o\right\}}^{}\sum_{n=0}^{\infty}\sum_{m=0}^{n}\les \Emn
\calB\smn\, r^{-(n+1)} \,  Y\smn(\theta,\phi)\ris \,,\quad r >a\,,
\end{equation}
is the perturbation potential.  If the object
were to be absent, Eq.~(\ref{eq4}) would hold for all $\#r$ with $\calB\smn\equiv 0$~$\forall \lec{s,m,n}\ric$.

\subsection{Source potential}\label{bvp-sp}
We proceed with the assumption that the coefficients $\calA\smn$ are
known but  the coefficients $\calB\smn$ are not. Furthermore, Eq.~(\ref{Phi-inc}) is required to hold
in some sufficiently large  open region that contains the spherical region $r\leq a$ but not the region containing the source
of $\Phi\inc(\#r)$.

Two illustrative examples of sources are a point charge $Q$ and a point dipole $\#p$.
Suppose, first, that the source potential is due to a point charge $Q$ located
at $\bro\equiv(\ro,\thetao,\phio)$ with $\ro>a$; then
\begin{equation}
\label{Phi-inc-ps}
\Phi\inc(\#r)=\frac{1}{4\pi\epso}\, \frac{Q}{\vert\#r-\bro\vert}\,.
\end{equation}
This potential can be expanded as \cite{Jackson}
\begin{equation}
\label{Phi-inc-ps-series}
\Phi\inc(\#r)=
\left\{
\begin{array}{c}
\displaystyle{
\sum_{s\in\left\{e,o\right\}}^{}\sum_{n=0}^{\infty}\sum_{m=0}^{n}\les\Emn
\bar{\calA}\smn   \,r^{-(n+1)}  \, Y\smn(\theta,\phi)\ris
},\,\,r>\ro,\\[8pt]
\displaystyle{
\sum_{s\in\left\{e,o\right\}}^{}\sum_{n=0}^{\infty}\sum_{m=0}^{n}\les \Emn
\calA\smn  \,r^n  \, Y\smn(\theta,\phi)\ris
},\,\,r<\ro,
\end{array}
\right.
\end{equation}
where the coefficients
\begin{subequations}
\begin{equation}
\label{Atilde-ps-p}
\bar{\calA}\smn=  \frac{Q}{\epso} \,\frac{1}{2n+1} \, \ro^n\,Y\smn(\thetao,\phio)
\end{equation}
and
\begin{equation}
\label{Atilde-ps-p}
\calA\smn = \frac{Q}{\epso} \,\frac{1}{2n+1} \,\ro^{-(n+1)}\,Y\smn(\thetao,\phio)\,.
\end{equation}
\end{subequations}

Suppose, next, that the source potential is due to a point dipole of moment  $\#p=p\hat{\#p}$  located
at $\bro\equiv(\ro,\thetao,\phio)$ with $\ro>a$; then  \cite{Tsitsas}
\begin{equation}
\label{Phi-inc-d}
\Phi\inc(\#r)=\frac{1}{4\pi\epso}\,\#p\.\nabla_{\rm o}\left(\frac{1}{\vert\#r-\bro\vert}\right)\,,
\end{equation}
where  $\nabla_{\rm o}(...)$ denotes the gradient with respect to $\bro$.
Equation~(\ref{Phi-inc-ps-series}) still holds, but with
\begin{subequations}
\begin{equation}
\label{Atilde-ps-d}
\bar{\calA}\smn=\frac{p}{\epso} \,\frac{1}{2n+1}\, \hat{\#p} \.\nabla_{\rm o}
\les\ro^n\,Y\smn(\thetao,\phio)\ris
\end{equation}
and
\begin{equation}
\label{Atilde-ps-d}
\calA\smn =\frac{p}{\epso} \,\frac{1}{2n+1}\, \hat{\#p}  \.\nabla_{\rm o}
\les\ro^{-(n+1)}\,Y\smn(\thetao,\phio)\ris\,.
\end{equation}
\end{subequations}

\subsection{Potential in the region $r<a$}\label{bvp-int}
Inside the dielectric sphere, the potential $\Phi(\#r)$ does not obey the Laplace equation;
instead,
\begin{equation}
\label{eq8}
\nabla\.\les\=A\.\=A\.\nabla\Phi(\#r)\ris=0\,.
\end{equation}
In order to solve this equation, let us make {an affine}  coordinate transformation:
\begin{equation}
\label{transf}
 \=A^{-1}\.\#r=\#r_q\equiv(r_q,\theta_q,\phi_q)\,,
 \end{equation}
where
\begin{subequations}
\begin{equation}
r_q = \vert  \=A^{-1}\.\#r\vert=r\sqrt{ \left(\alphax^2\cos^2\phi+\alphay^2\sin^2\phi\right)\sin^2\theta+\cos^2\theta }\,\geq0\,,
\end{equation}
\begin{equation}
\theta_q=\cos^{-1}\left(\frac{r}{\vert\=A^{-1}\.\#r\vert} \cos\theta\right)\,,
\end{equation}
and
\begin{equation}
\phi_q=\tan^{-1}\left(\frac{\alphay}{\alphax}\tan\phi\right)\,.
\end{equation}
\end{subequations}
The bijective transformation (\ref{transf}) maps a sphere into an ellipsoid
since $\alphax>0$ and $\alphay>0$, with
$\theta_q$ lying in the same quadrant as $\theta$ and $\phi_q$ in the same
quadrant as $\phi$.

Then, Eq.~(\ref{eq8}) can be written as
\begin{equation}
\label{eq9}
\nabla_q\.\les \nabla_q\Phi(\#r_q)\ris=0\,,
\end{equation}
i.e.,
\begin{equation}
\label{eq9b}
  \nabla_q^2\,\Phi(\#r_q) =0\,,
\end{equation}
which is the Laplace equation in the transformed space.
Its solution is given by \cite{Jackson,Moon}
\begin{equation}
\label{eq10}
\Phi(\#r_q)= \sum_{s\in\left\{e,o\right\}}^{}\sum_{n=0}^{\infty}\sum_{m=0}^{n}\lec \Emn \les
\calC\smn\, r_q^n + \calD\smn\, r_q^{-(n+1)}\ris\, Y\smn(\theta_q,\phi_q)\ric\,.
\end{equation}
We must set $\calD\smn\equiv0$ in order
to exclude terms on the right side
of Eq.~(\ref{eq10}) that are not regular at the origin. Thereafter, on inverting the coordinate transformation,
we obtain the internal  potential
\begin{equation}
\label{Phi-int}
\Phi\ind(\#r)= \sum_{s\in\left\{e,o\right\}}^{}\sum_{n=0}^{\infty}\sum_{m=0}^{n}\les \Emn
\calC\smn\,  Z\smn(\#r)\ris\,,\quad r<a\,,
\end{equation}
where
\begin{equation}
\left.\begin{array}{l}
Z\emn(\#r) = \vert\=A^{-1}\.\#r\vert^n\,P_n^m \les\frac{r}{\vert\=A^{-1}\.\#r\vert} \cos\theta\ris
\cos\les m \tan^{-1}\left(\frac{\alphay}{\alphax}\,\tan\phi\right)\ris
\\[8pt]
Z\omn(\#r) = \vert\=A^{-1}\.\#r\vert^n\,P_n^m \les\frac{r}{\vert\=A^{-1}\.\#r\vert} \cos\theta\ris
\sin\les m \tan^{-1}\left(\frac{\alphay}{\alphax}\,\tan\phi\right)\ris
\end{array}\right\}\,.
\end{equation}

\subsection{Boundary conditions}\label{bvp-bc}
Since the tangential component of the electric field must be continuous across the interface $r=a$, and
as there is no reason for the electric field to have an infinite magnitude anywhere on that interface, the
potential must be continuous across that interface; hence,
\begin{equation}
\label{bc1}
\Phi\inc(r,\theta,\phi) + \Phi\sca(r,\theta,\phi)=\Phi\ind(r,\theta,\phi)\,,
\quad r=a\,,\quad
\theta\in[0,\pi]\,,\quad \phi\in[0,2\pi)\,.
\end{equation}
Likewise, with the assumption of the interface $r=a$ being charge-free, the normal component of the
electric displacement must be continuous across that interface; hence,
\begin{equation}
\label{bc2}
\frac{\partial}{\partial r}\les\Phi\inc(r,\theta,\phi) + \Phi\sca(r,\theta,\phi)\ris=
\epsr\ur\.\=A\.\=A\.\nabla
\Phi\ind(r,\theta,\phi)\,,
\quad r=a\,,\quad
\theta\in[0,\pi]\,,\quad \phi\in[0,2\pi)\,,
\end{equation}
where $\ur=(\ux\cos\phi+\uy\sin\phi)\sin\theta+\uz\cos\theta$.

\subsection{Transition matrix}\label{bvp-me}

After (i) substituting Eqs.~(\ref{Phi-inc}), (\ref{Phi-sca}), and (\ref{Phi-int}) in Eq.~(\ref{bc1}), (ii) then
multiplying both sides of the resulting equation by $Y\smnp(\theta,\phi)\sin\theta$, and (iii) finally integrating
over  $\theta\in[0,\pi]$ and $\phi\in[0,2\pi)$, we get
\begin{equation}
\label{mateq1}
\sum\smn \les
\calA\smn\calI\double+\calB\smn\calJ\double\ris
= \sum\smn \les \calC\smn\calK\double\ris\,,
\end{equation}
where
\begin{subequations}

\begin{equation}
\label{def-I}
\calI\double =a^n\,\delta_{ss^\prime}
\delta_{mm^\prime}\delta_{nn^\prime} \,,
\end{equation}
\begin{equation}
\label{def-J}
\calJ\double =a^{-(n+1)} \,
\delta_{ss^\prime}
\delta_{mm^\prime}\delta_{nn^\prime} \,,
\end{equation}
and
\begin{equation}
\label{def-K}
\calK\double = \Emn \int_{\phi=0}^{2\pi}
\int_{\theta=0}^{\pi} \, Z\smn(a,\theta,\phi)Y\smnp(\theta,\phi)\sin\theta\,\mathrm{d}\theta\,\mathrm{d}\phi
 \,.
\end{equation}
\end{subequations}
Similarly, after (i) substituting Eqs.~(\ref{Phi-inc}), (\ref{Phi-sca}), and (\ref{Phi-int}) in Eq.~(\ref{bc2}), (ii) then
multiplying both sides of the resulting equation by $Y\smnp(\theta,\phi)\sin\theta$, and (iii) finally integrating
over  $\theta\in[0,\pi]$ and $\phi\in[0,2\pi)$, we get
\begin{equation}
\label{mateq2}
\sum\smn \les
\calA\smn\calL\double+\calB\smn\calM\double\ris
= \sum\smn \les \calC\smn\calN\double\ris\,,
\end{equation}
where
\begin{subequations}

\begin{equation}
\label{def-L}
\calL\double =na^{n-1}\,
 \delta_{ss^\prime}
\delta_{mm^\prime}\delta_{nn^\prime} \,,
\end{equation}
\begin{equation}
\label{def-M}
\calM\double =-(n+1)a^{-(n+2)}\,
 \delta_{ss^\prime}
\delta_{mm^\prime}\delta_{nn^\prime} \,,
\end{equation}
and
\begin{equation}
\label{def-N}
\calN\double =\epsr\,\Emn\int_{\phi=0}^{2\pi}
\int_{\theta=0}^{\pi} \, \ur\.\=A\.\=A\.
\lec\les\nabla
Z\smn(\#r)\ris\Big\vert_{r=a}\ric
Y\smnp(\theta,\phi)\sin\theta\,\mathrm{d}\theta\,\mathrm{d}\phi
 \,,
\end{equation}

\end{subequations}

After  truncating the indexes $n$  and $n^\prime$ so that only $n\in[0,N]$ and $n^\prime\in[0,N]$
are considered with $N>0$,
Eqs.~(\ref{mateq1}) and (\ref{mateq2}) can be put together  in matrix form symbolically as
\begin{equation}
\les\begin{array}{cc} \calI & \calJ\\ \calL & \calM\end{array}\ris
\les\begin{array}{c} \calA\\ \calB  \end{array}\ris
= \les\begin{array}{c} \calK\\ \calN  \end{array}\ris\,\les\calC\ris\,,
\end{equation}
which leads to the solution
\begin{equation}
\les\begin{array}{c} \calC\\ \calB  \end{array}\ris=
\les\begin{array}{cc} \calK & -\calJ\\ \calN & -\calM\end{array}\ris^{-1}
\les\begin{array}{c} \calI\\ \calL  \end{array}\ris\,\les\calA\ris\,.
\end{equation}
Thus, the coefficients $\calB\smn$ and $\calC\smn$ can be determined in terms of the coefficients $\calA\smn$.

The perturbational characteristics of the  {anisotropic}  dielectric
sphere are encapsulated in the transition matrix $\calT$ that relates the column vectors
$\calB$ and $\calA$ via
\begin{equation} \label{T-def}
 \calB =\calT\,\calA\,,
 \end{equation}
where
\begin{equation} \label{Tmat}
\calT= - \calJ^{-1} \left(\calM\,\calJ^{-1}-\calN\,\calK^{-1}\right)^{-1}\,
\left(\calL\,\calI^{-1}-\calN\,\calK^{-1}\right)\,\calI\,.
\end{equation}
The transition matrix  is a diagonal matrix when the sphere is made of an isotropic dielectric material
(i.e., $\=A=\=I$) because $\calN\,\calK^{-1}$ is then a diagonal matrix. In general, $\calN\,\calK^{-1}$    is not a diagonal   matrix
when the sphere is made of an
 {anisotropic}  dielectric material,  so that $\calT$ is
not a diagonal matrix either.

\subsection{Reduction of computational effort}\label{bvp-rce}
Computational effort for the integrals (\ref{def-K}) and  (\ref{def-N}) can be significantly reduced on noting
that
\begin{equation}
\left.\begin{array}{ll}
\sin[m(\pi+\phi)]=(-)^m\,\sin(m\phi)\,,\quad & \sin[m(\pi+\phi_q)]=(-)^m\,\sin(m\phi_q)\\[5pt]
\cos[m(\pi+\phi)]=(-)^m\,\cos(m\phi)\,,\quad & \cos[m(\pi+\phi_q)]=(-)^m\,\cos(m\phi_q)\\[5pt]
P_n^m[\cos(\pi-\theta)]=(-)^{n+m} \,P_n^m(\cos \theta) \,,\quad &
P_n^m[\cos(\pi-\theta_q)]=(-)^{n+m} \,P_n^m(\cos \theta_q)
\end{array}\right\}\,;
\end{equation}
furthermore, $\theta_q$ lies in the same quadrant as $\theta$ and $\phi_q$ in the same
quadrant as $\phi$. Therefore,
\begin{eqnarray}
\nonumber
\calK\double &=&  \les 1+(-)^{m+m^\prime}\ris \les 1+(-)^{m+m^\prime+n+n^\prime}\ris\Emn
\\[7pt]
&&
\label{def-K-symm}
\times\int_{\phi=0}^{\pi}
\int_{\theta=0}^{\pi/2} \, Z\smn(a,\theta,\phi)Y\smnp(\theta,\phi)\sin\theta\,\mathrm{d}\theta\,\mathrm{d}\phi
 \,
\end{eqnarray}
and
\begin{eqnarray}
\nonumber
\calN\double &=& \les 1+(-)^{m+m^\prime}\ris \les 1+(-)^{m+m^\prime+n+n^\prime}\ris\epsr\, \Emn
\\[7pt]
\label{def-N-symm}
&&\times
\int_{\phi=0}^{\pi}
\int_{\theta=0}^{\pi/2} \, \ur\.\=A\.\=A\.
\lec\les\nabla
Z\smn(\#r)\ris\Big\vert_{r=a}\ric
Y\smnp(\theta,\phi)\sin\theta\,\mathrm{d}\theta\,\mathrm{d}\phi
 \,
\end{eqnarray}
can be used instead of Eqs.~(\ref{def-K}) and (\ref{def-N}), respectively.

 \subsection{Symmetries of the transition matrix}\label{bvp-symm}

By virtue of its definition through Eq.~(\ref{T-def}), the transition matrix $\calT$ does not depend on the source potential.
This matrix depends only on the radius $a$ and the constitutive parameters $\alphax$, $\alphay$, and $\epsr$ of
the perturbing sphere.

Let us denote each element of the transition matrix defined in Eq. (\ref{Tmat}) by $\calT\double$. It was verified numerically that $\calT\double\ne0$     if the following three conditions are satisfied:
\begin{itemize}
\item[(i)] $s=s^\prime$,
\item[(ii)] $m$ and $m^\prime$ have the same parity (i.e., even or odd), and
\item[(iii)] $n$ and $n^\prime$ have the same parity.
\end{itemize}

Finally, let the transition matrix elements be denoted as $\calT\double\one$ for a specific choice $\lec\alphax,\alphay\ric$
of the anisotropy parameters, but as $\calT\double\two$ after $\alphax$ and $\alphay$ have been interchanged without
changing $\epsr$. In other words,  $\calT\double\one$ changes to $\calT\double\two$ when the sphere is rotated about
the $z$ axis by $\pi/2$.
Then,  the following relationships exist between the pre- and post-rotation transition matrixes:
\begin{itemize}
\item $\calT_{\rm ss mm nn^\prime}\one=\calT_{\rm ss mm nn^\prime}\two$ when $m$ is even,
\item $\calT_{\rm ee mm  nn^\prime}\one= \calT_{\rm oo mm  nn^\prime}\two$
and $\calT_{\rm oo mm  nn^\prime}\one= \calT_{\rm ee mm  nn^\prime}\two$ when $m$ is odd,
\item  $\calT_{\rm ss mm^\prime nn^\prime}\one=-\calT_{\rm ss mm^\prime nn^\prime}\two$
when $m \ne m^\prime$ and both are even, and
\item $\calT_{\rm ee mm^\prime  nn^\prime}\one= -\calT_{\rm oo mm^\prime  nn^\prime}\two$
and $\calT_{\rm oo mm^\prime  nn^\prime}\one=-\calT_{\rm ee mm^\prime  nn^\prime}\two$ when $m \ne m^\prime$ and both are odd.
\end{itemize}

\subsection{Asymptotic expression for perturbation potential}\label{bvp-asymp}
Equation~(\ref{Phi-sca}) can be written as
\begin{eqnarray}
\nonumber
\Phi\sca(\#r)
&=& \frac{1}{4\pi} \lec
r^{-1}\calB_{\rm e00}
+ 3 r^{-2}\les
\calB_{\rm e01} \cos\theta +2 \left(\calB_{\rm e11}\cos\phi + \calB_{\rm o11}\sin\phi\right)\sin\theta
\ris\ric
\\[5pt]
\label{Phi-sca-1}
&&+
\sum_{s\in\left\{e,o\right\}}^{}\sum_{n=2}^{\infty}\sum_{m=0}^{n}\les \Emn
\calB\smn\, r^{-(n+1)} \,  Y\smn(\theta,\phi)\ris \,,\quad r >a\,.
\end{eqnarray}
Since $\calB_{\rm e00}=0$ emerges from calculations for a sphere whether $\=A=\=I$ or not,
the asymptotic behavior of the perturbation potential far away from the sphere is given by
\begin{equation}
\Phi\sca(r, \theta, \phi)=\frac{\fsca(\theta,\phi)}{r^2}+\mathcal{O}\left(\frac{1}{r^3}\right),\,\,\,r\rightarrow \infty\,,
\end{equation}
where the asymptotic perturbation
\begin{equation}
\label{fsca-def}
\fsca(\theta,\phi)=
\frac{3}{4\pi}\les\calB_{\rm e01} \cos\theta +2 \left(\calB_{\rm e11}\cos\phi + \calB_{\rm o11}\sin\phi\right)\sin\theta\ris\,.
\end{equation}
Accordingly, the first term on the right side of Eq.~(\ref{fsca-def}) does not exist in the equatorial plane (i.e., $\theta=\pi/2$)
whereas the second term is absent on the $z$ axis (i.e., $\theta\in\lec0,\pi\ric$).

Now, the perturbation potential $\Phi\sca(r, \theta, \phi)$ must depend on
 the   source as well as on the radius $a$ of the sphere.
For the two sources chosen for illustrative results, $\fsca$ must depend linearly on
both the sign and the magnitude of $Q$ or $p$ (as appropriate).
 The location of either of the two sources   enters the potential expressions only by means of the
source-potential coefficients $\mathcal{A}\smn$, which are proportional to $\ro^{-(n+1)}$ for the
point charge and to $\ro^{-(n+2)}$ for the
point dipole, according to
Eqs. (\ref{Atilde-ps-p}) and (\ref{Atilde-ps-d}). Since the transition matrix $\calT$ does not depend on the
source, the perturbation-potential coefficients $\calB_{\rm e01}$, $\calB_{\rm e11}$, and $\calB_{\rm o11}$   increase/decrease
as $\ro$ decreases/increases. Accordingly, the magnitude of $\fsca$   increases/decreases
as $\ro$ decreases/increases.

Furthermore, from Eqs.~(\ref{def-I}), (\ref{def-J}), (\ref{def-L}), and (\ref{def-M}), we get
\begin{equation}
\left.\begin{array}{ll}
\calI\double\propto {a}^{n}\,,  &\calJ\double\propto {a}^{-(n+1)}
\\[5pt]
\calL\double\propto {a}^{n-1}\,,  &\calM\double\propto {a}^{-(n+2)}
\end{array}
\right\}\,.
\end{equation}
Since the coefficients $\calA\smn$ cannot depend on $a$, it follows then from Eqs.~(\ref{mateq1}) and (\ref{mateq2})
that
\begin{equation}
\label{eq39}
\calB\smn\propto {a}^{2n+1}\,.
\end{equation}
Dimensional analysis of Eq.~(\ref{eq4}) also supports this proportionality. Equations~(\ref{fsca-def}) and (\ref{eq39}) then yield
accordingly,
\begin{equation}
\fsca(\theta,\phi)  \propto a^3\,.
\end{equation}

\section{Numerical Results and Discussion}\label{nrd}

\subsection{Preliminaries}\label{nrd1}

For all numerical results presented in this section, we fixed
\begin{equation}
\epsr = \frac{3 \, \eps_\text{ave}}{\alphax^{-2}+ \alphay^{-2} + 1},
\end{equation}
and $\eps_\text{ave}=3$.
Note that $\epsr=\eps_\text{ave}$ when the sphere is made of an isotropic material (i.e., $\=A=\=I$).

A Mathematica\texttrademark~program was written to calculate the normalized functions
\begin{subequations}
\begin{equation}
\label{Phisca-def}
\tilde{\Phi}_\text{pert}(r,\theta,\phi)=
\displaystyle{
\rtilde^2\, \Phi_\text{pert}(r,\theta,\phi)
}\,
\end{equation}
and
\begin{equation}
\label{Fsca-def}
\Fsca(\theta,\phi)=
\displaystyle{
\frac{\fsca(\theta,\phi)}{\pi\,a^2\,{\Phi\inc^{\rm ref}}}
}\,,
\end{equation}
\end{subequations}
where the normalized radius $\rtilde=r/a$ and
 $\Phi\inc^{\rm ref} \ne 0$ is some reference value of the source potential. The reference value can be chosen as
 $ \Phi\inc^{\rm ref}=\Phi\inc(\#0) =  \calA_{e00}/4\pi$. This choice works well if the
source is an external point charge and also if the source is an external point dipole
except when $\#p\.\bro=0$.
By virtue of its definition and the dependencies of $\fsca$ discussed in Sec.~\ref{bvp-asymp}, $\Fsca$
\begin{itemize}
\item[(i)] is independent of $Q$ or $p$ (as appropriate),
\item[(ii)] increases/decreases as $\ro$ decreases/increases, and
\item[(iii)] is directly proportional to $a$.
\end{itemize}

 A convergence test was carried out with respect to $N$, by calculating the integral
\begin{equation}
I(\rtilde)=
\int_{\phi=0}^{2 \pi}
\int_{\theta=0}^{\pi} \tilde{\Phi}_\text{pert}^2(\rtilde\,a,\theta,\phi) \sin \theta \, \mathrm{d} \theta \, \mathrm{d} \phi,
\end{equation}
at diverse values of $\rtilde>1$ as $N$ was incremented by unity. The iterative process of increasing $N$ was terminated when
$I(\rtilde)$ within a preset tolerance of $1\%$.  The adequate value of $N$ was higher for  lower $\rtilde$, {with $N=5$ sufficient for $\rtilde\geq 5$}.

The theory described in Sec.~\ref{bvp} was validated by comparing its results for the perturbation of the
source potential  by an isotropic dielectric sphere   with the corresponding exact solutions available in the literature. First, the source was taken to be a point charge located on the $+z$~axis (i.e. $\thetao=0$) at $\ro=10a$; note that $\phio$ is irrelevant when $\sin\thetao=0$.
Excellent agreement was obtained with respect to the exact solution \cite{Stratton} for all examined values of $a$ and $\epsr$.
Next, the point charge was replaced by a point dipole. Again, excellent agreement was found   with respect to the corresponding exact solutions \cite{Tsitsas,Zurita}.

\subsection{Normalized asymptotic perturbation $\Fsca(\theta, \phi)$}\label{nrd2}
Having clarified   in Sec.~\ref{nrd1} the {effects} of the parameters $Q$, $p$, $\ro$, and $a$ on the normalized asymptotic perturbation $\Fsca(\theta, \phi)$, we present next numerical results for the variations of $\Fsca(\theta, \phi)$ as a function of the sphere's anisotropy parameters $\alphax$ and $\alphay$ for the following three cases:
\begin{itemize}
\item[]Case 1: $\alphax \neq \alphay = 1$,
\item[]Case 2: $\alphay\neq\alphax=1$, and
\item[]Case 3:  $\alphax = \alphay \neq 1$.
\end{itemize}
Plots of the perturbation-potential coefficients $\calB_{\rm e01}$, $\calB_{\rm e11}$, and $\calB_{\rm o11}$ as functions of $\alphax$ and $\alphay$ are examined   in conjunction with the corresponding plots of   $\Fsca(\theta, \phi)$ versus $\theta$ and $\phi$ for
 $a = 5$~cm, $\ro = 2 a$, $\thetao = \pi/4$, and $\phio = \pi/3$. {All  calculations were made for either a point charge of magnitude $Q=10^{-10}$~C or a point dipole of moment $p=10^{-10}$~C~m.}

\subsubsection{Case 1 ($\alphax =\balpha$, $\alphay = 1$)}

\begin{figure}[H]
 \centering
\includegraphics[width=0.4\linewidth]{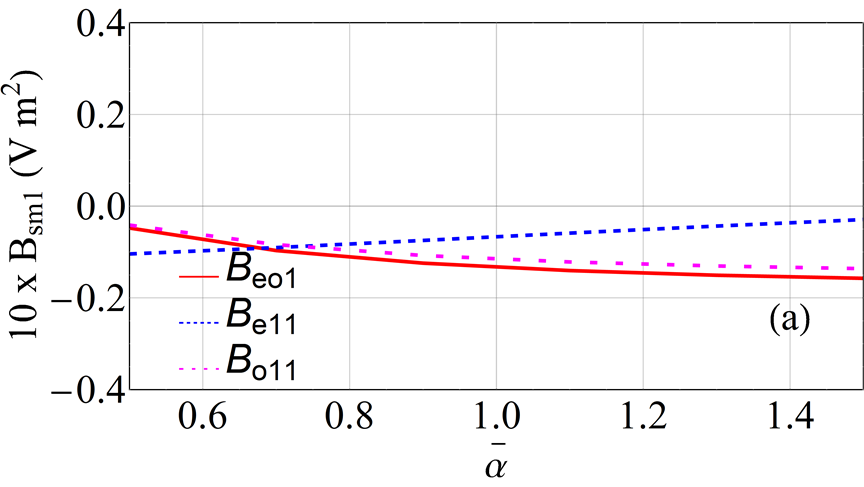}
\includegraphics[width=0.4\linewidth]{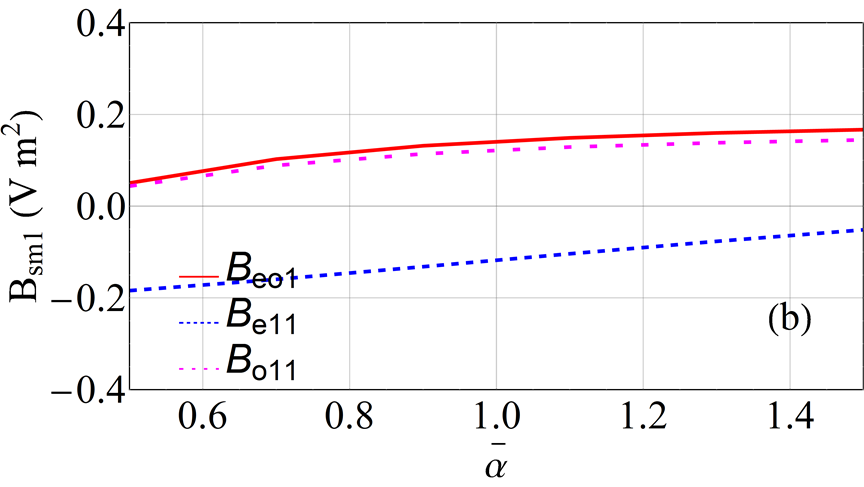} \\
\includegraphics[width=0.4\linewidth]{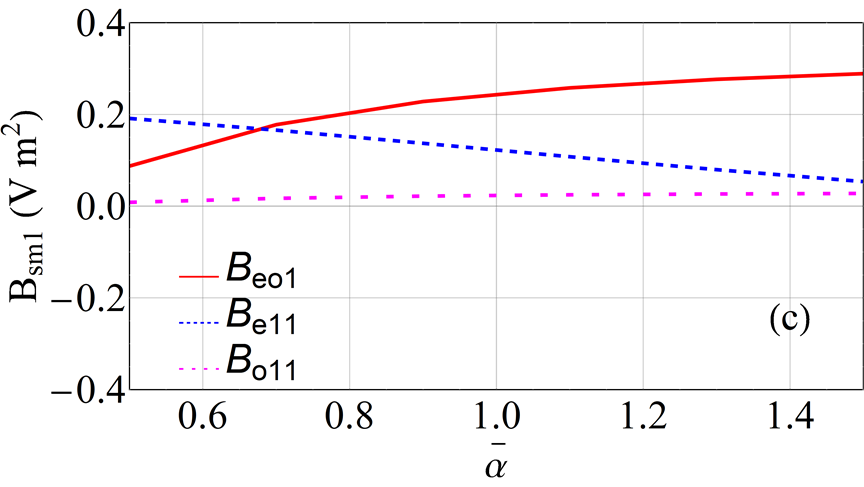}
\includegraphics[width=0.4\linewidth]{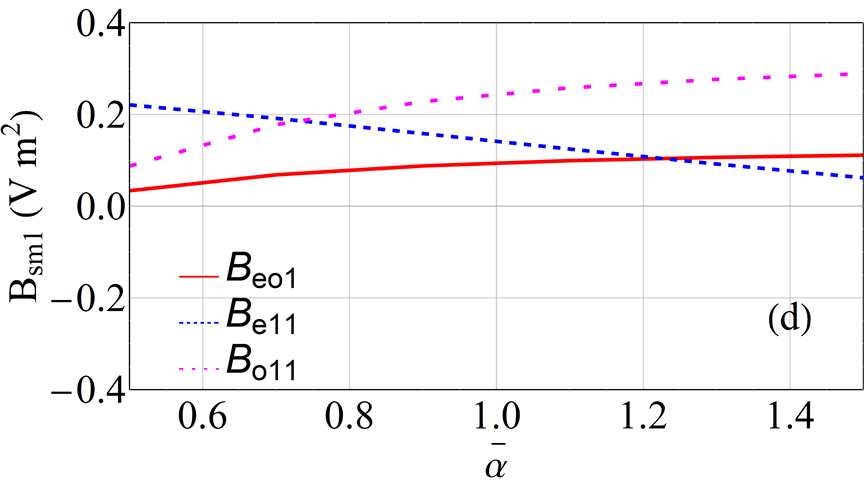}
\caption{$\calB_{\rm e01}$, $\calB_{\rm e11}$, and $\calB_{\rm o11}$ vs $\balpha \in [0.5,1.5]$ when
$\alphax=\balpha$, $\alphay=1$, 
and the source is either
 (a) a point charge     or (b-d) a point dipole    with (b) $\up = \ux$,
(c) $\up = \uy$, and (d) $\up = \uz$, respectively.}
\label{c1}
\end{figure}

\begin{figure}[H]
 \centering
\includegraphics[width=0.2\linewidth]{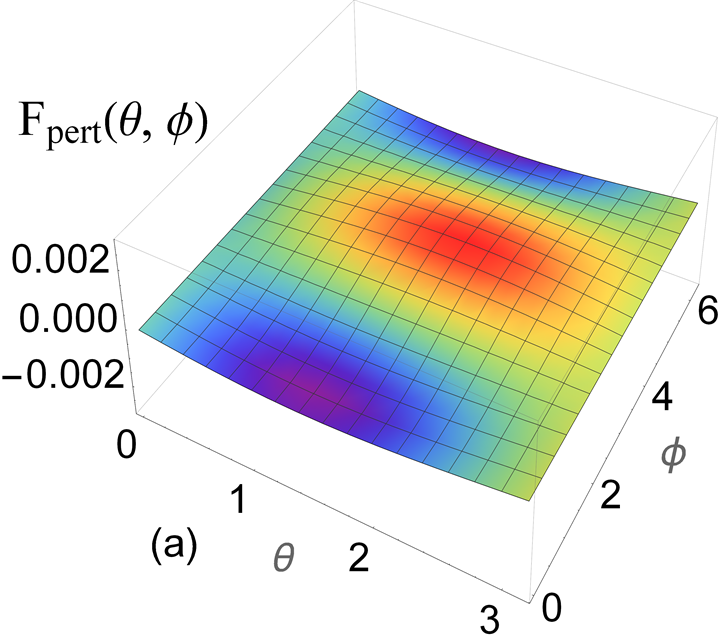}
\includegraphics[width=0.2\linewidth]{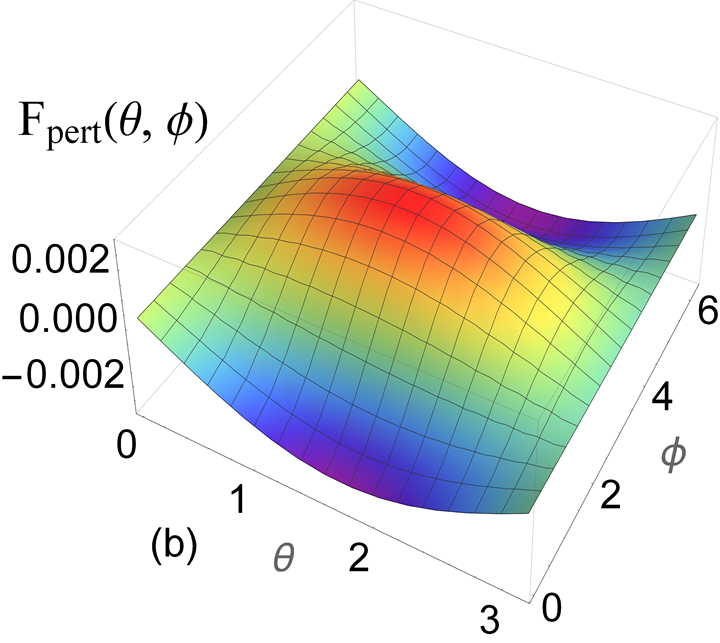} \\
\includegraphics[width=0.2\linewidth]{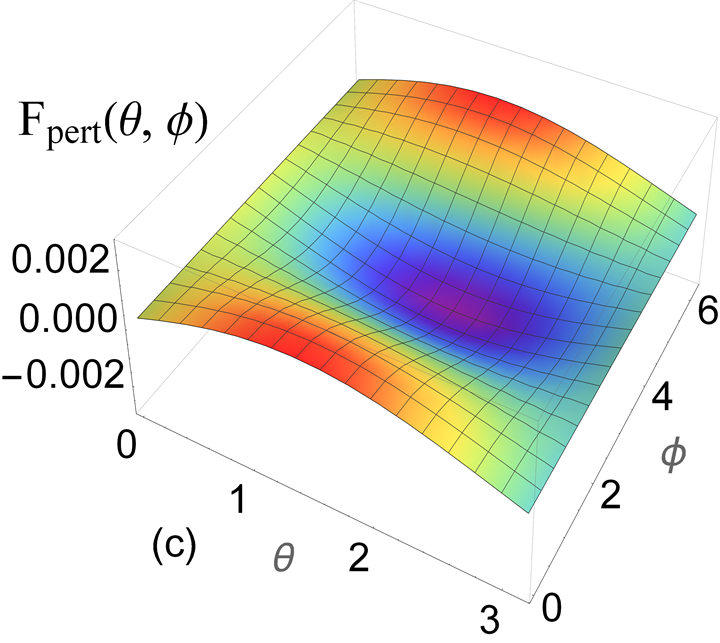}
\includegraphics[width=0.2\linewidth]{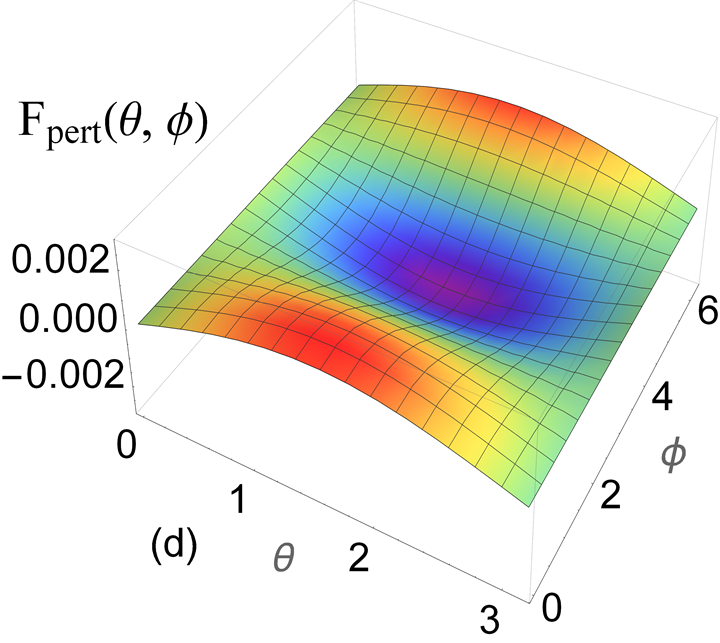}
\caption{$\Fsca(\theta, \phi)$ vs $\theta$ and $\phi$ when $\alphax = 0.5$, $\alphay = 1$, 
and the source is either
 (a) a point charge     or (b-d) a point dipole     with (b) $\up = \ux$,
(c) $\up = \uy$, and (d) $\up = \uz$, respectively.}
\label{c2}
\end{figure}

\begin{figure}[H]
 \centering
\includegraphics[width=0.2\linewidth]{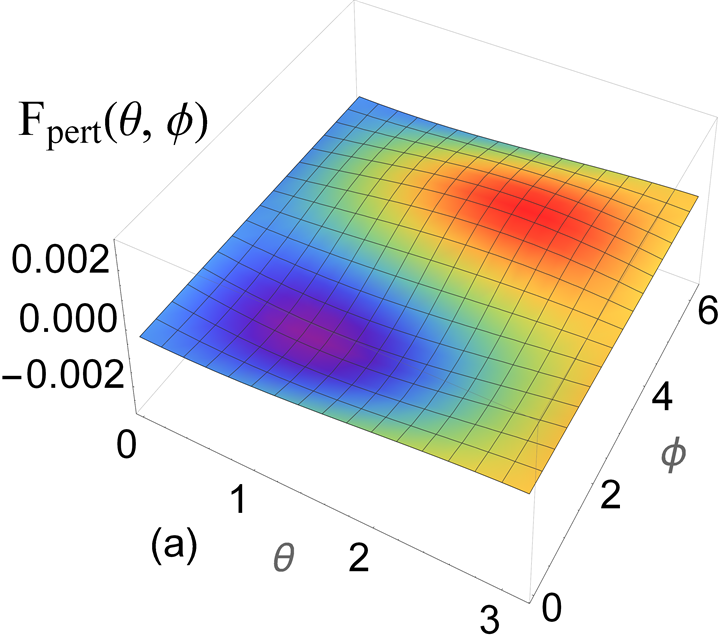}
\includegraphics[width=0.2\linewidth]{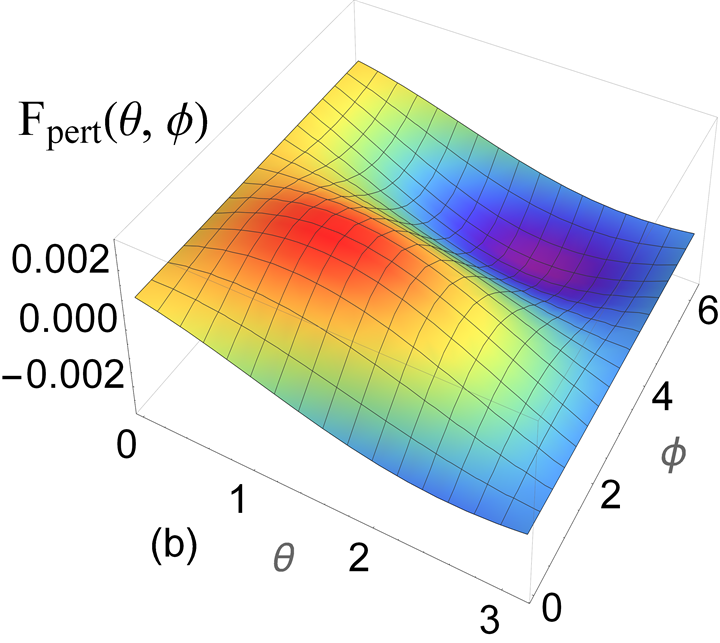} \\
\includegraphics[width=0.2\linewidth]{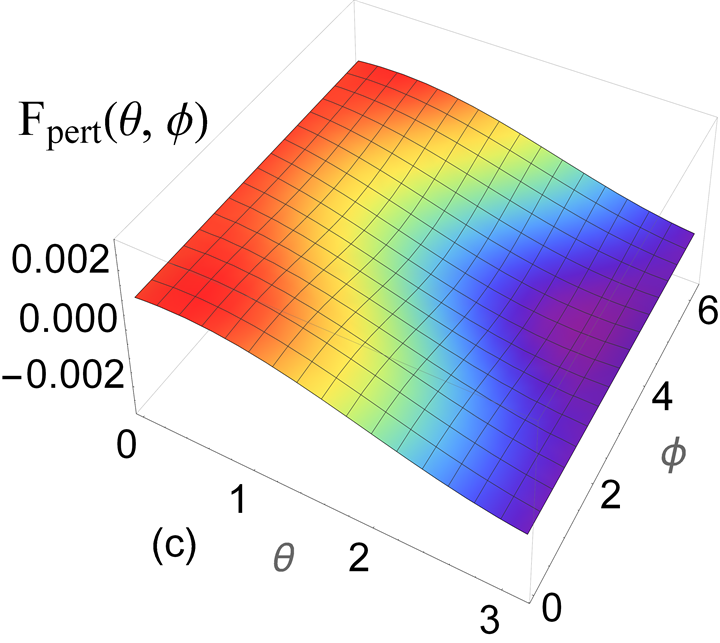}
\includegraphics[width=0.2\linewidth]{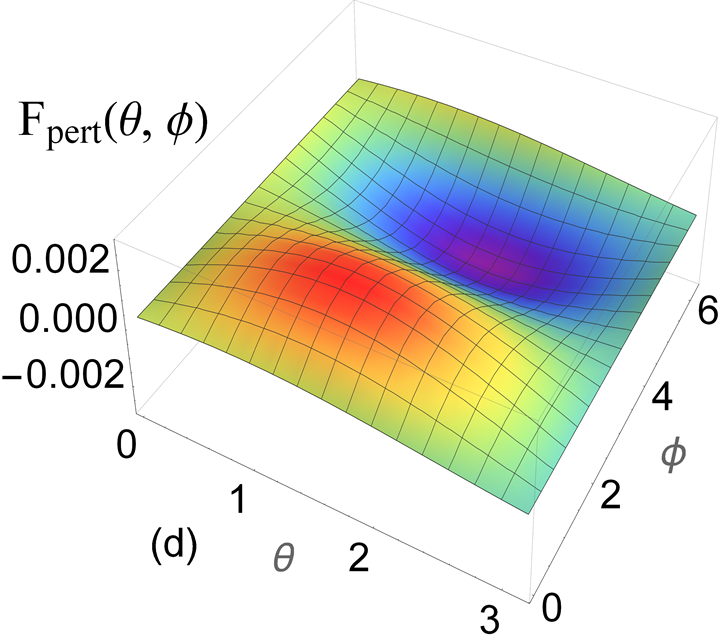}
\caption{As in Fig.~\ref{c2}, except for $\alphax=1.5$.}
\label{c3}
\end{figure}

We varied $\alphax =\balpha \in [0.5,1.5]$ but kept $\alphay = 1$ fixed. Figure \ref{c1} shows plots of $\calB_{\rm e01}$, $\calB_{\rm e11}$, and $\calB_{\rm o11}$ versus $\balpha$ for a point-charge  source as well as for a point-dipole source with $\up \in\lec \ux,  \uy, \uz\ric$. Angular profiles of the  $\Fsca(\theta, \phi)$ for the same  sources are depicted in Fig.~\ref{c2} for $\balpha =0.5$, and in Fig.~\ref{c3} for $\balpha =1.5$.

For a point-charge source, $\calB_{\rm e01}$ decreases with $\balpha \in [0.5,1.5]$ in  Fig.~\ref{c1}(a).
Thus, $\Fsca(0,\phi)$ decreases but $\Fsca(\pi,\phi)$ increases as $\alphax$ changes from $0.5$ to $1.5$, as can be gathered
from Figs.~\ref{c2}(a) and \ref{c3}(a).
Also, $\calB_{\rm e11}$ increases and $\calB_{\rm o11}$ decreases with increasing $\alphax$. Therefore, $\Fsca(\pi/2,0)$ increases  but $\Fsca(\pi/2,\pi/2)$ decreases as $\alphax$ changes from $0.5$ to $1.5$.

Next, for the  point-dipole sources, $\calB_{\rm e01}$ increases with increasing $\balpha \in [0.5,1.5]$ for all three dipole orientations, as is clear from   Figs.~\ref{c1}(b)--(d); the largest increase is observed for $\up = \uy$. Hence, a comparison of Figs.~\ref{c2}(b)--(d)
and  \ref{c3}(b)--(d) reveals that
$\Fsca(0,\phi)$ increases but $\Fsca(\pi,\phi)$ decreases as $\alphax$ changes from $0.5$ to $1.5$.   The rate of these increases or decreases is highest for  $\up = \uy$, moderate for $\up = \ux$, and lowest for $\up = \uz$.

Besides, for $\up = \ux$, both $\calB_{\rm e11}$ and $\calB_{\rm o11}$ increase with $\alphax$ in Fig. \ref{c1}(b) and, thus, $\Fsca(\pi/2,0)$ and $\Fsca(\pi/2,\pi/2)$ also increase with $\alphax$ in Figs. \ref{c2}(b) and \ref{c3}(b). On the other hand, for $\up = \uy$ and $\up = \uz$, $\calB_{\rm e11}$ decreases in Figs. \ref{c1}(c)  but $\calB_{\rm o11}$ increases   in Fig. \ref{c1}(d) as  $\alphax$ increases. Therefore, $\Fsca(\pi/2,0)$ decreases and $\Fsca(\pi/2,\pi/2)$ increases with $\alphax$, as can be gathered from comparing
Figs. \ref{c2}(c) and (d) with   Figs. \ref{c3}(c) and  (d), respectively.

\subsubsection{Case 2 ($\alphax=1$, $\alphay=\balpha$)}

\begin{figure}[H]
 \centering
\includegraphics[width=0.3\linewidth]{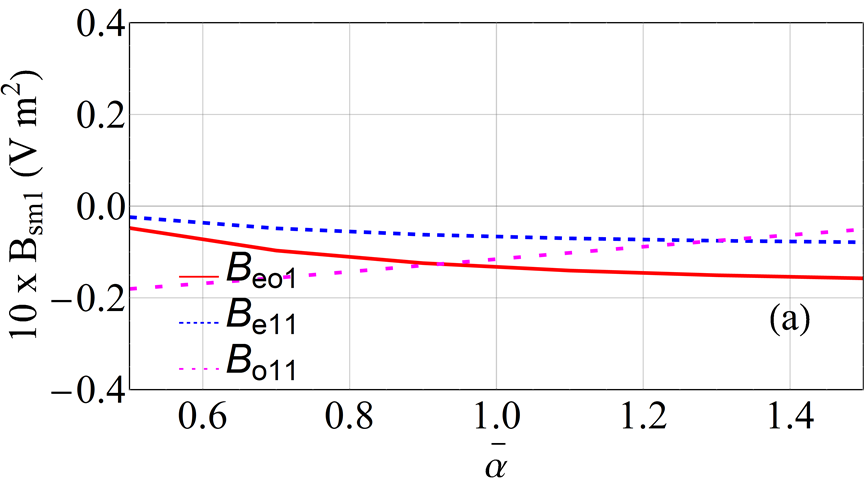}
\includegraphics[width=0.3\linewidth]{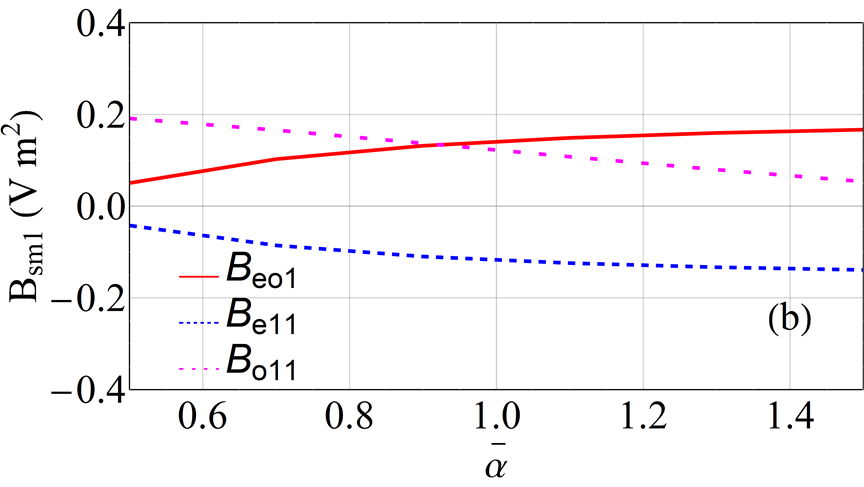} \\
\includegraphics[width=0.3\linewidth]{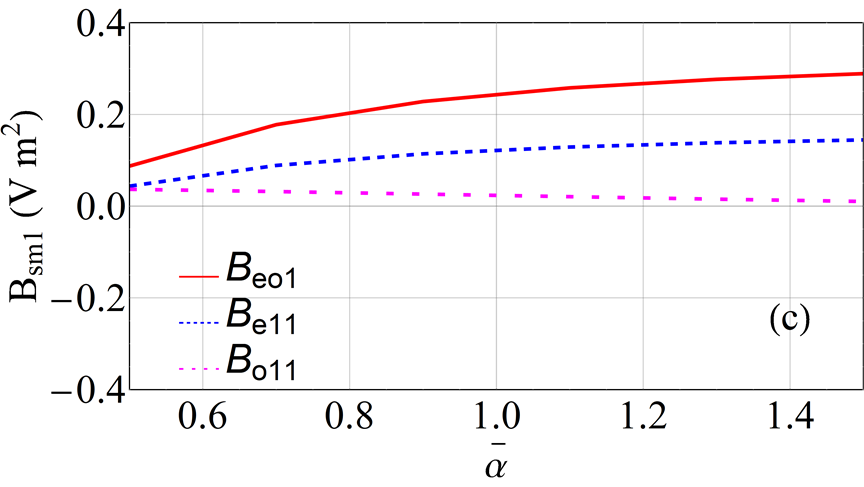}
\includegraphics[width=0.3\linewidth]{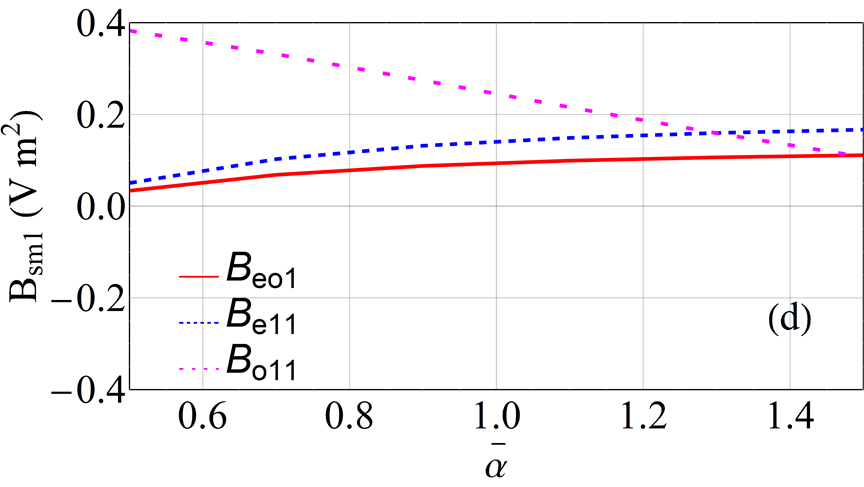}
\caption{$\calB_{\rm e01}$, $\calB_{\rm e11}$, and $\calB_{\rm o11}$ vs $\balpha \in [0.5,1.5]$ when $\alphax=1$, $\alphay=\balpha$,  
and the source is either
 (a) a point charge     or (b-d) a point dipole     with (b) $\up = \ux$,
(c) $\up = \uy$, and (d) $\up = \uz$, respectively.}
\label{c4}
\end{figure}

\begin{figure}[H]
 \centering
\includegraphics[width=0.2\linewidth]{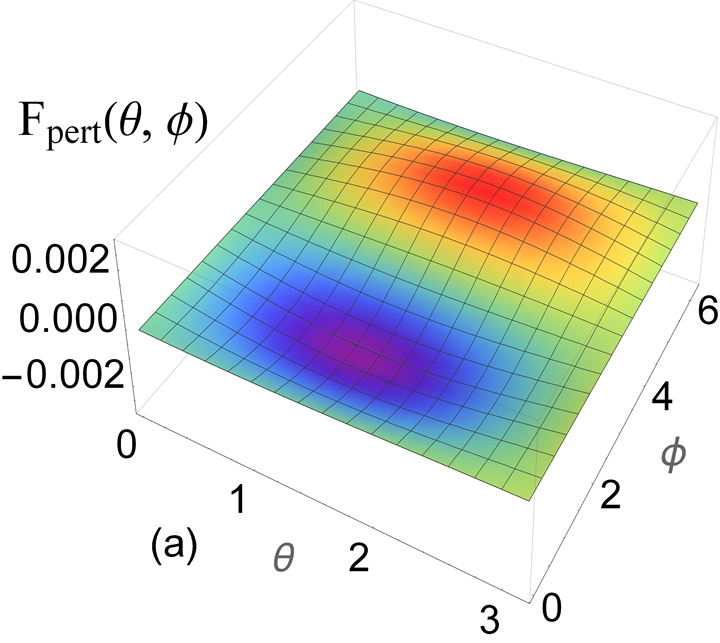}
\includegraphics[width=0.2\linewidth]{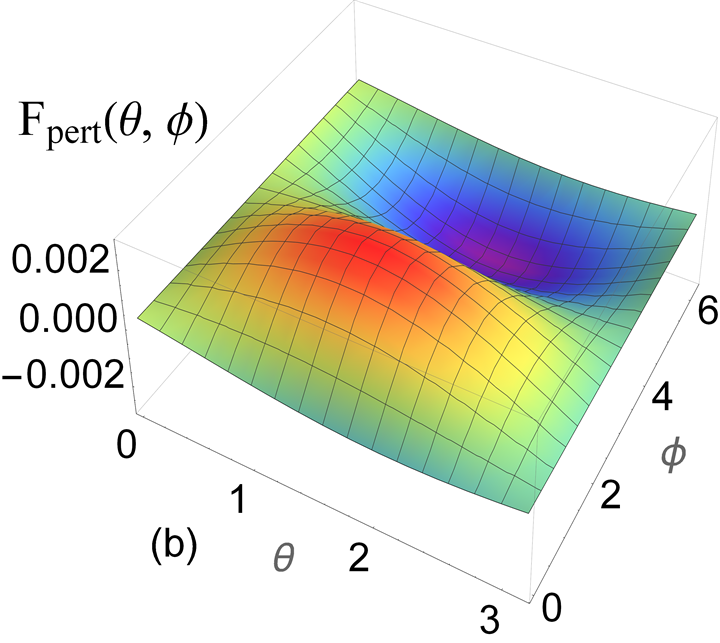} \\
\includegraphics[width=0.2\linewidth]{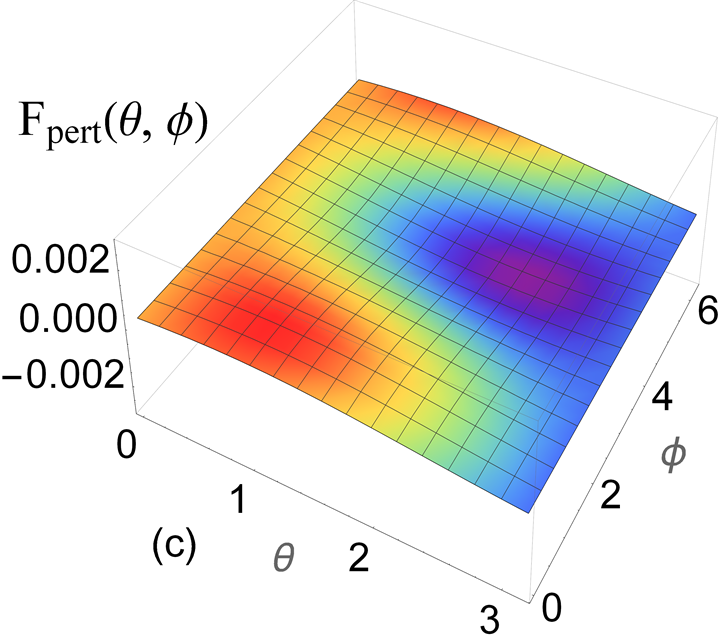}
\includegraphics[width=0.2\linewidth]{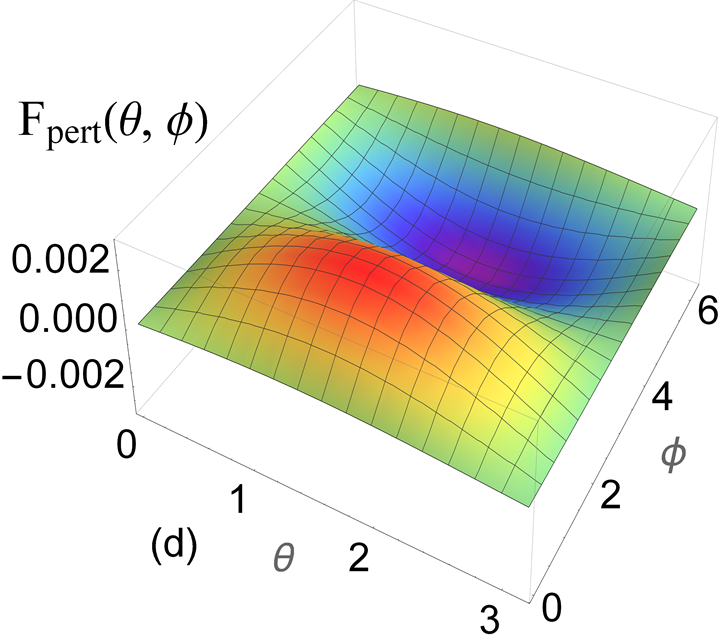}
\caption{As in Fig. \ref{c2}, except for $\alphax=1$ and $\alphay = 0.5$.}
\label{c5}
\end{figure}

\begin{figure}[H]
 \centering
\includegraphics[width=0.2\linewidth]{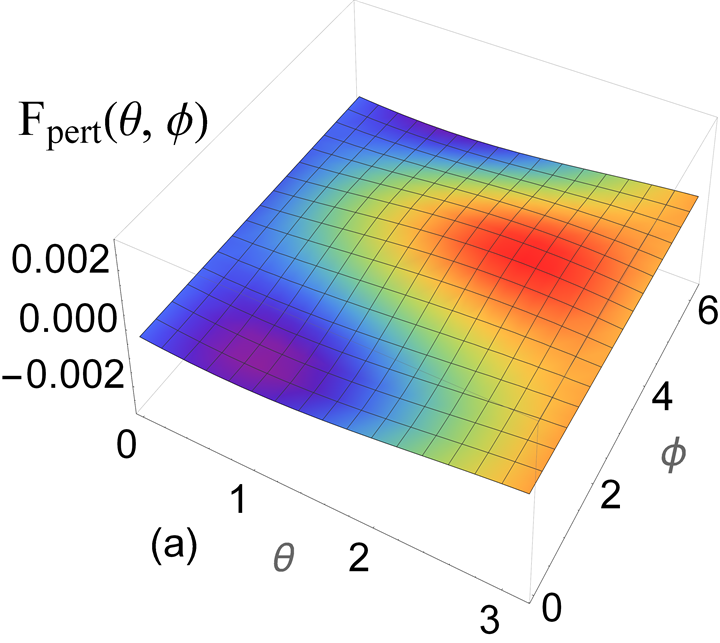}
\includegraphics[width=0.2\linewidth]{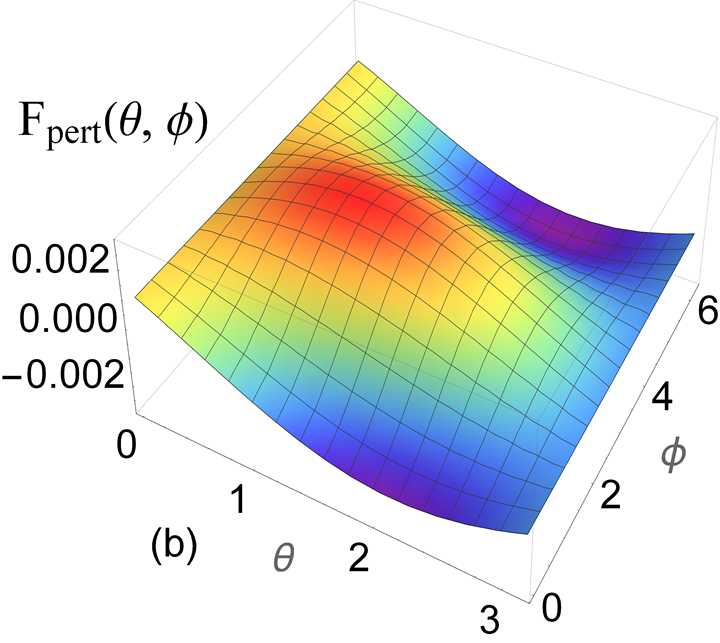} \\
\includegraphics[width=0.2\linewidth]{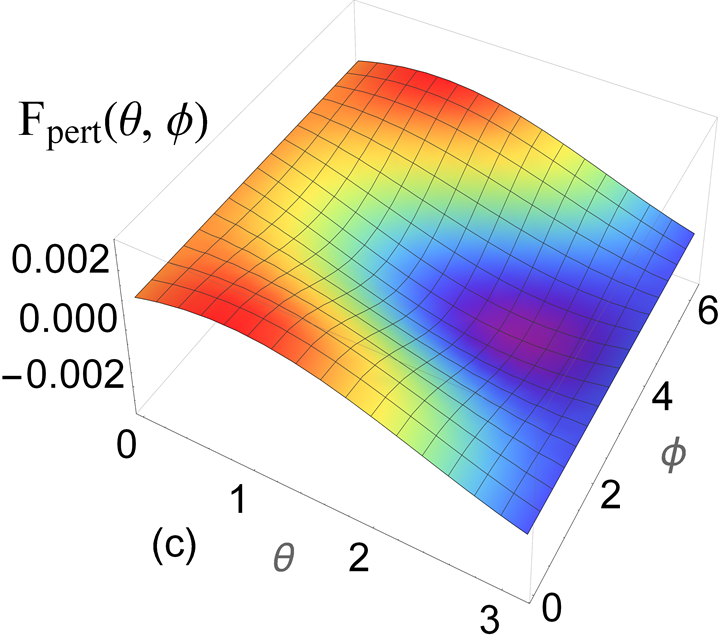}
\includegraphics[width=0.2\linewidth]{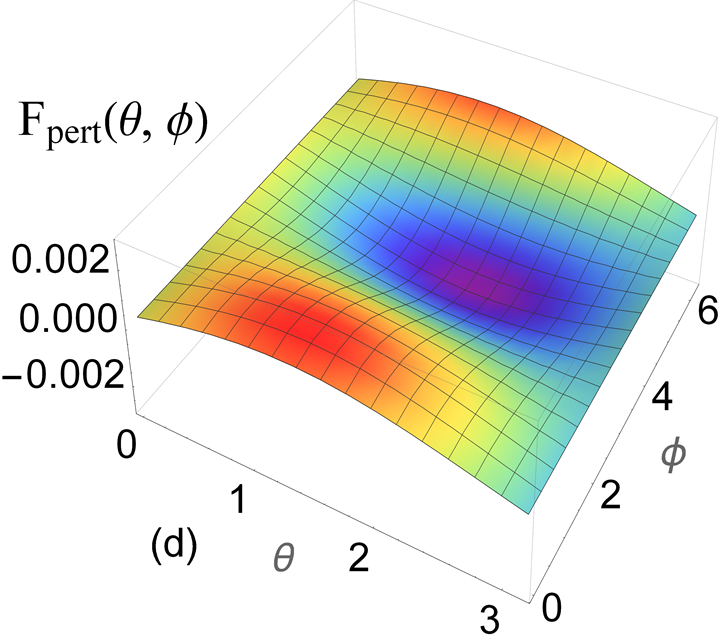}
\caption{As in Fig. \ref{c5}, except for $\alphay=1.5$.}
\label{c6}
\end{figure}

Next, we fixed $\alphax = 1$ but varied $\alphay \in[0.5,1.5]$. The dependencies of the coefficients $\calB_{\rm e01}$, $\calB_{\rm e11}$, and $\calB_{\rm o11}$ on $\alphay$ are depicted in Fig.~\ref{c4}, whereas the angular profiles of $\Fsca(\theta, \phi)$ are depicted in Fig.~\ref{c5} for $\alphay=0.5$ and   Fig.~\ref{c6} for $\alphay=1.5$.

For all four types of   source considered,
$\calB_{\rm e01}$ varies with $\balpha$ in Case 1
in the same way as it varies with $\balpha$ in Case 2.  Hence, the characteristics of $\Fsca(0, \phi)$ and $\Fsca(\pi, \phi)$ in Case 2 replicate those in Case 1.
Also, if $\calB_{\rm e11}$ or $\calB_{\rm o11}$ is an increasing (decreasing) function of $\balpha$ in Case 1, then it is a decreasing (increasing) function of $\balpha$ in Case 2.  Therefore, the characteristics of $\Fsca(\pi/2, 0)$ and $\Fsca(\pi/2, \pi/2)$ in Case 2 are opposed to those in Case 1.

\subsubsection{Case 3 ($\alphax = \alphay =\balpha$)}

\begin{figure}[H]
 \centering
\includegraphics[width=0.3\linewidth]{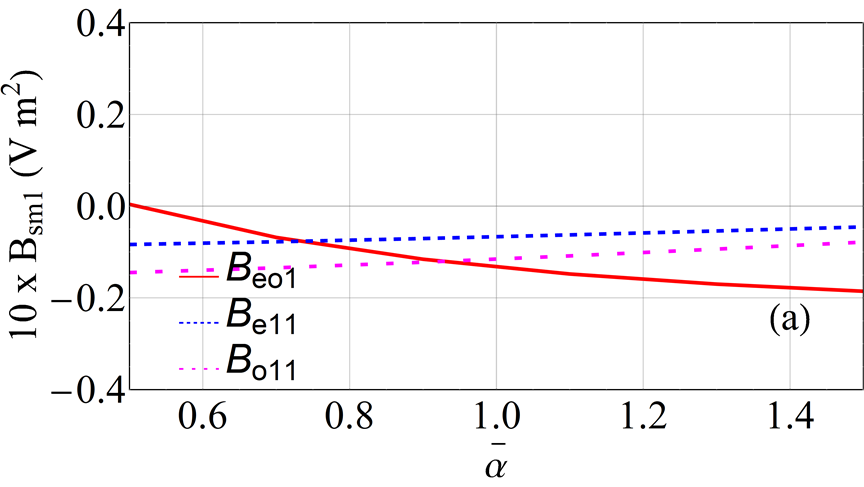}
\includegraphics[width=0.3\linewidth]{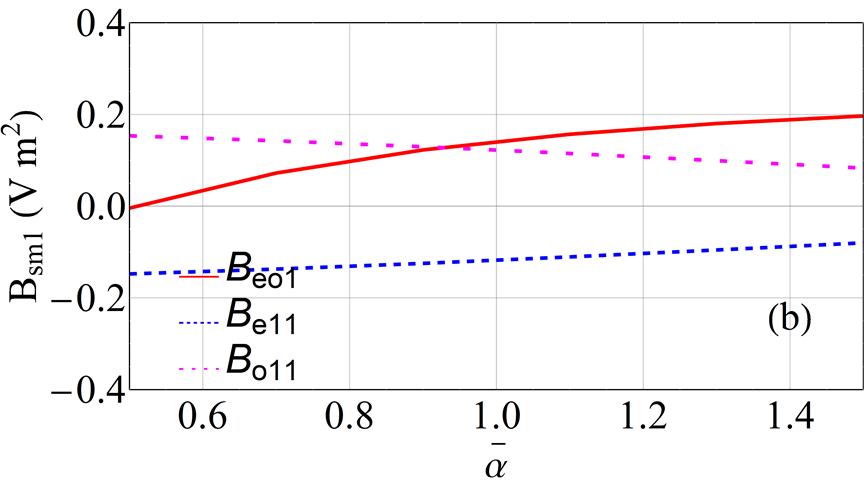} \\
\includegraphics[width=0.3\linewidth]{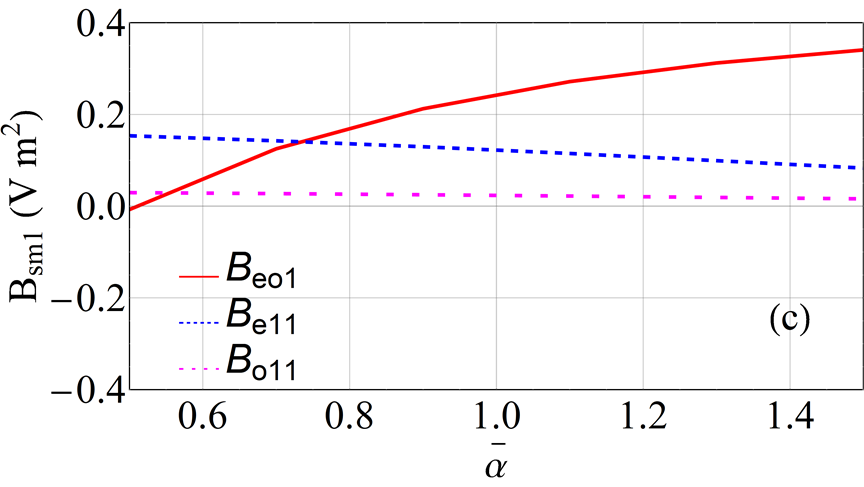}
\includegraphics[width=0.3\linewidth]{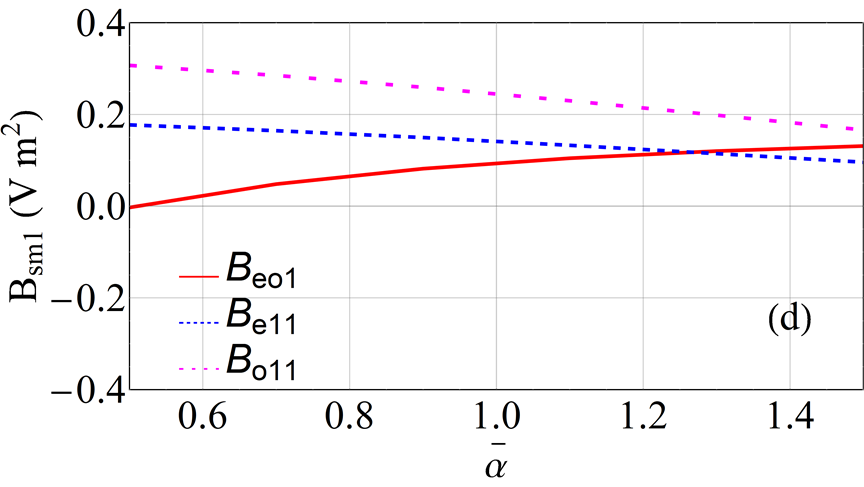}
\caption{As in Fig. \ref{c1}, except for $\alphax = \alphay = \balpha$.}
\label{c7}
\end{figure}

\begin{figure}[H]
 \centering
\includegraphics[width=0.2\linewidth]{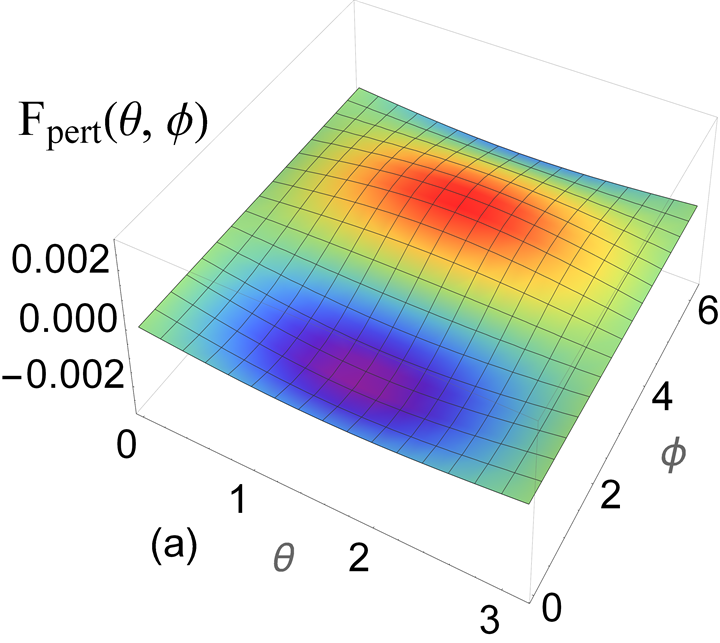}
\includegraphics[width=0.2\linewidth]{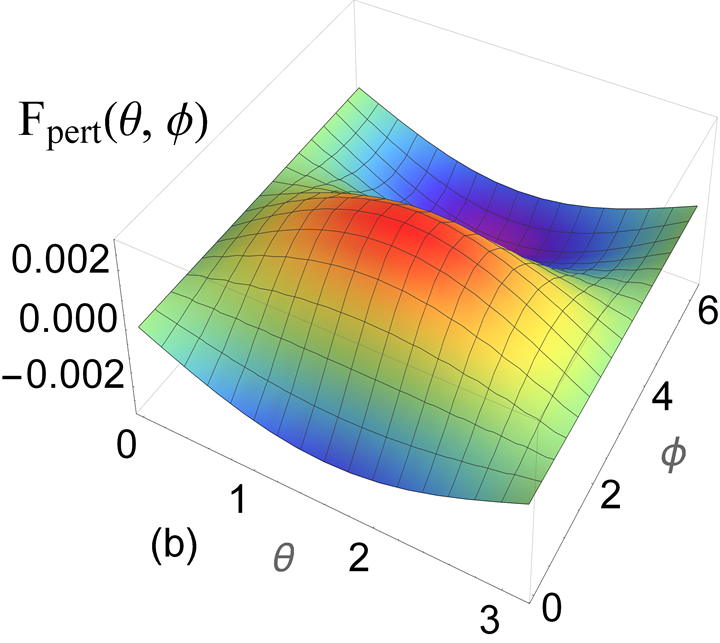} \\
\includegraphics[width=0.2\linewidth]{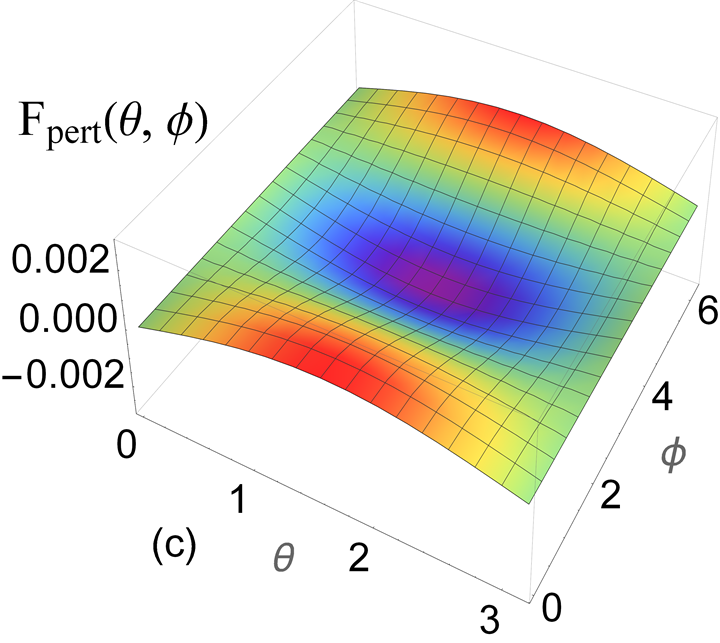}
\includegraphics[width=0.2\linewidth]{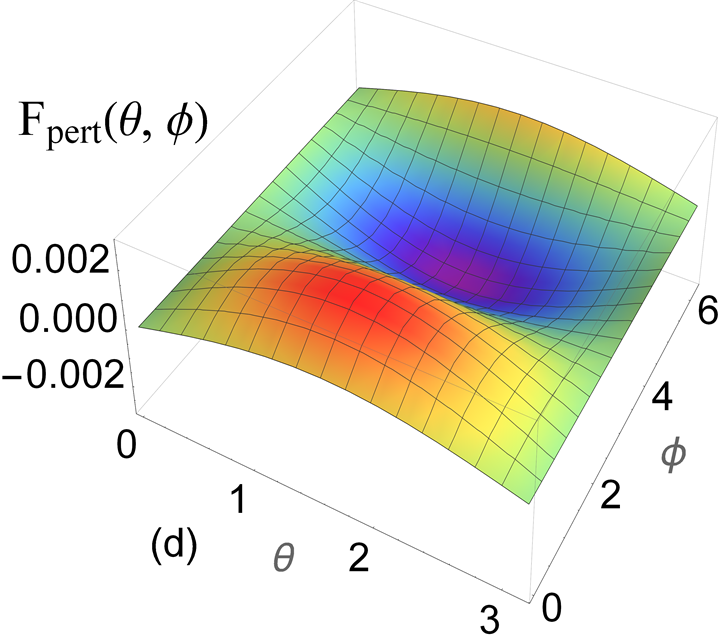}
\caption{As in Fig. \ref{c2}, except for $\alphax=\alphay = 0.5$.}
\label{c8}
\end{figure}

\begin{figure}[H]
 \centering
\includegraphics[width=0.2\linewidth]{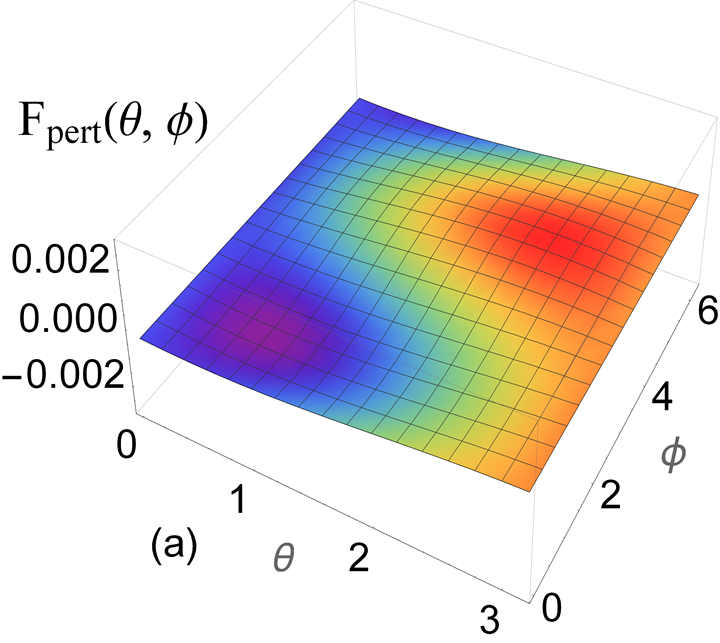}
\includegraphics[width=0.2\linewidth]{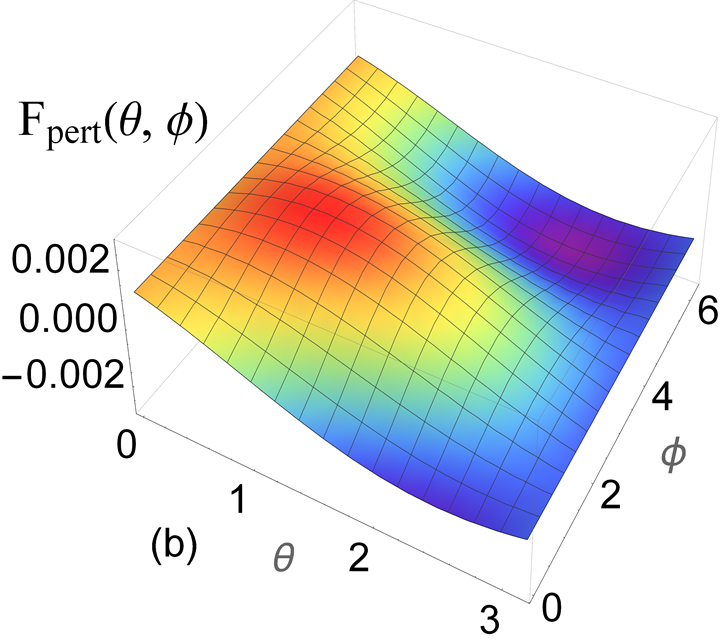} \\
\includegraphics[width=0.2\linewidth]{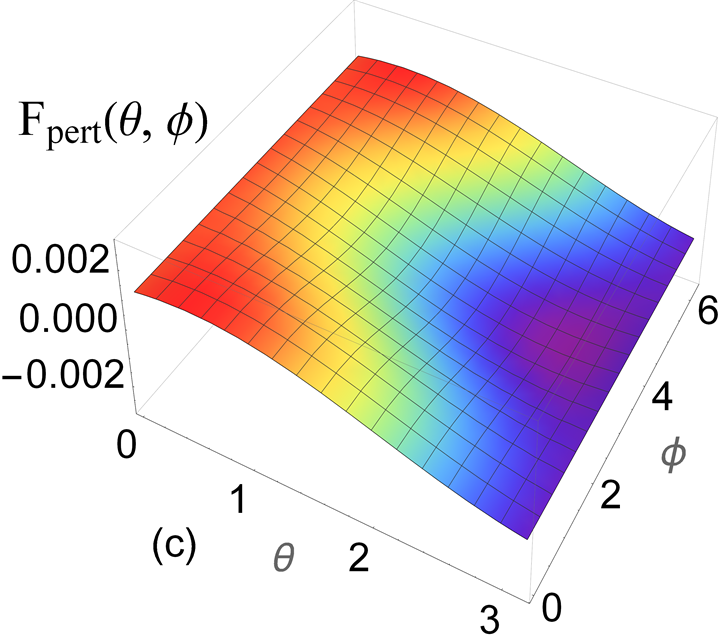}
\includegraphics[width=0.2\linewidth]{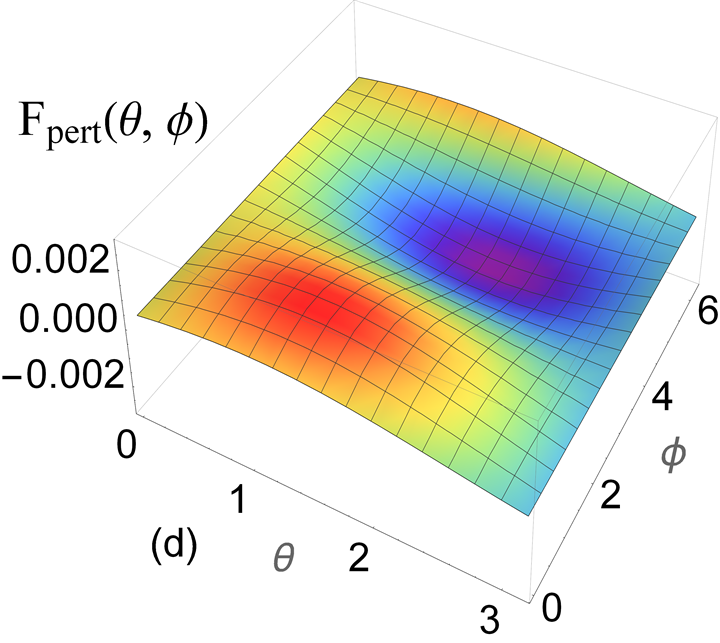}
\caption{As in Fig. \ref{c8}, except for $\alphax=\alphay = 1.5$.}
\label{c9}
\end{figure}

Finally, we set $\alphax = \alphay = \balpha$.
The corresponding plots for $\calB\smn$ vs $\balpha \in [0.5,1.5]$ are depicted in Figure \ref{c7}, and for $\Fsca(\theta, \phi)$ vs $\theta$ and $\phi$ are depicted in Fig.~\ref{c8} for $\balpha =0.5$, and in Fig.~\ref{c9} for $\balpha =1.5$.

The curves of $\calB_{\rm e01}$ vs $\balpha \in [0.5,1.5]$ have the same increasing/decreasing tendencies with the respective ones in Cases 1 and 2; however, the values of $\calB_{\rm e01}$ and the rate of increase/decrease w.r.t. $\balpha$ are definitely different. Furthermore, in Case 3, the increasing/decreasing tendencies of $\calB_{\rm e11}$ and $\calB_{\rm o11}$ with $\balpha$ are as those in Cases 1 and 2, respectively.

\subsection{Normalized perturbation potential $\tilde{\Phi}_\text{pert}(r,\theta,\phi)$}

\begin{figure}[H]
 \centering
\includegraphics[width=0.2\linewidth]{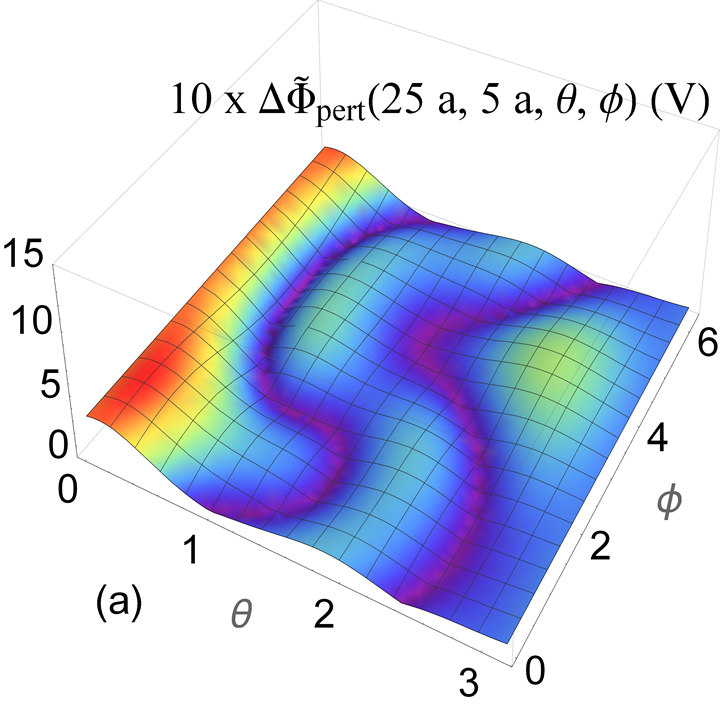} 
\includegraphics[width=0.2\linewidth]{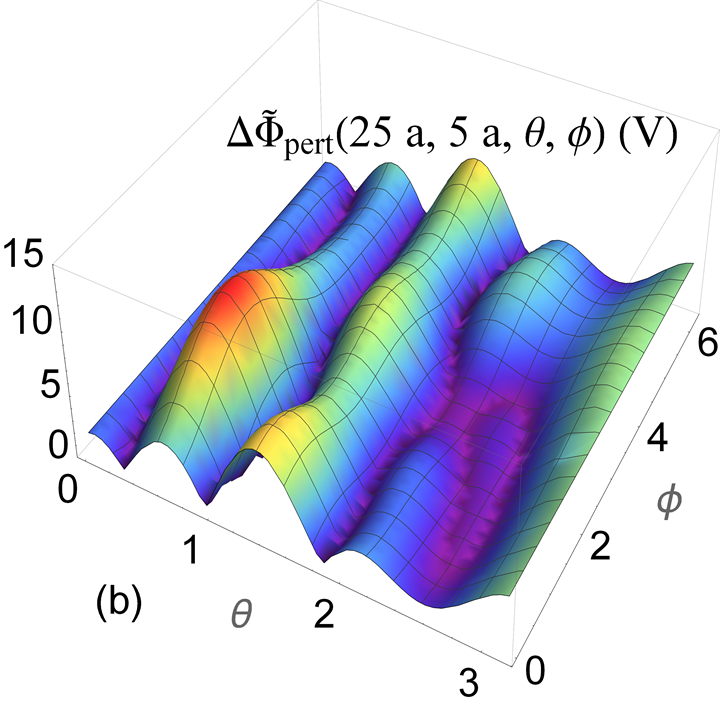} \\
\includegraphics[width=0.2\linewidth]{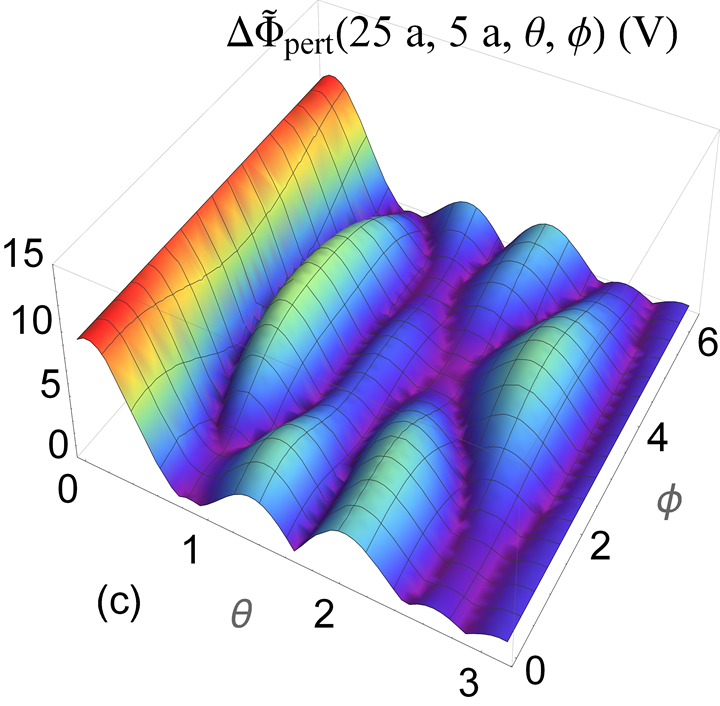} 
\includegraphics[width=0.2\linewidth]{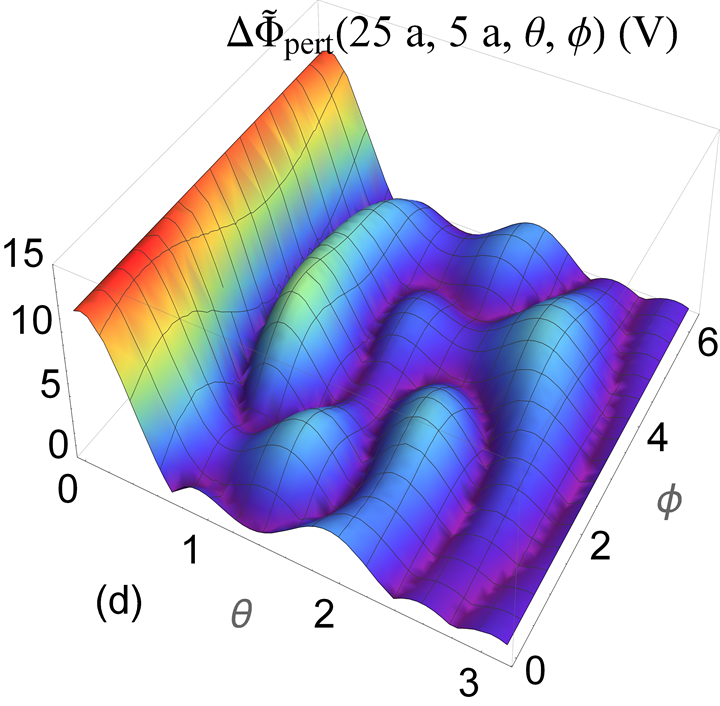}
\caption{$\Delta\tilde{\Phi}_\text{pert}(25a,5a,\theta,\phi)$  versus $\theta$ and $\phi$ when
$\alphax = 1.2$,   $\alphay = 1.6$, 
and the source is either
 (a) a point charge     or (b-d) a point dipole     with (b) $\up = \ux$, with (b) $\up = \ux$,
(c) $\up = \uy$, and (d) $\up = \uz$, respectively.} 
\label{c10}
\end{figure}

\begin{figure}[H]
 \centering
\includegraphics[width=0.2\linewidth]{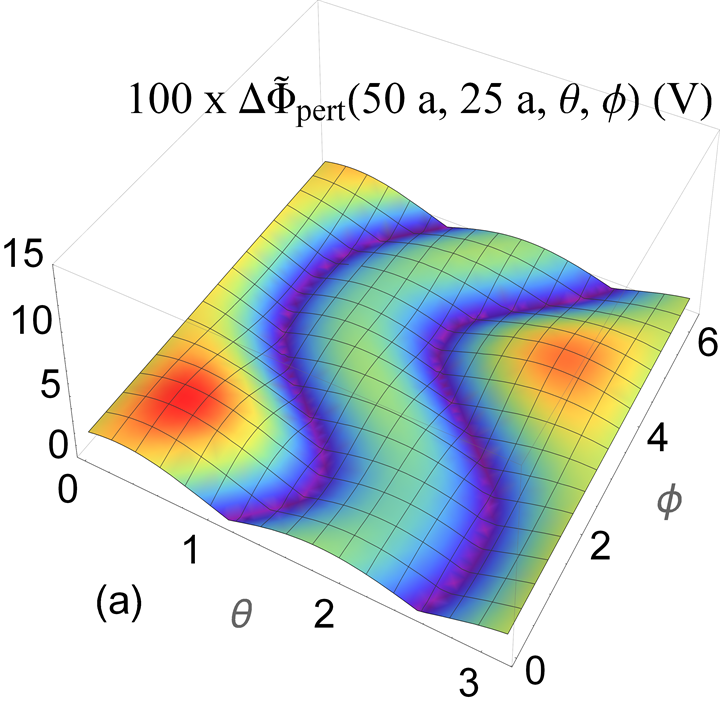} 
\includegraphics[width=0.2\linewidth]{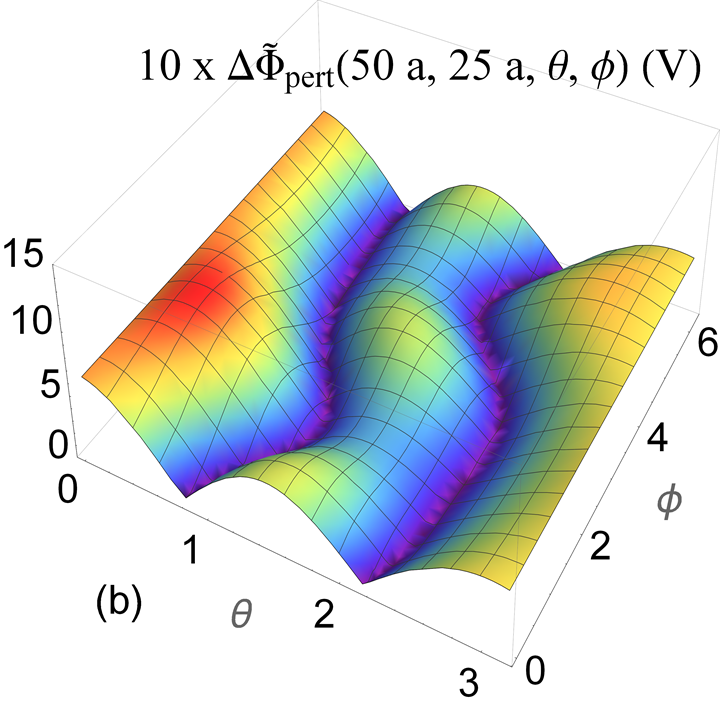} \\
\includegraphics[width=0.2\linewidth]{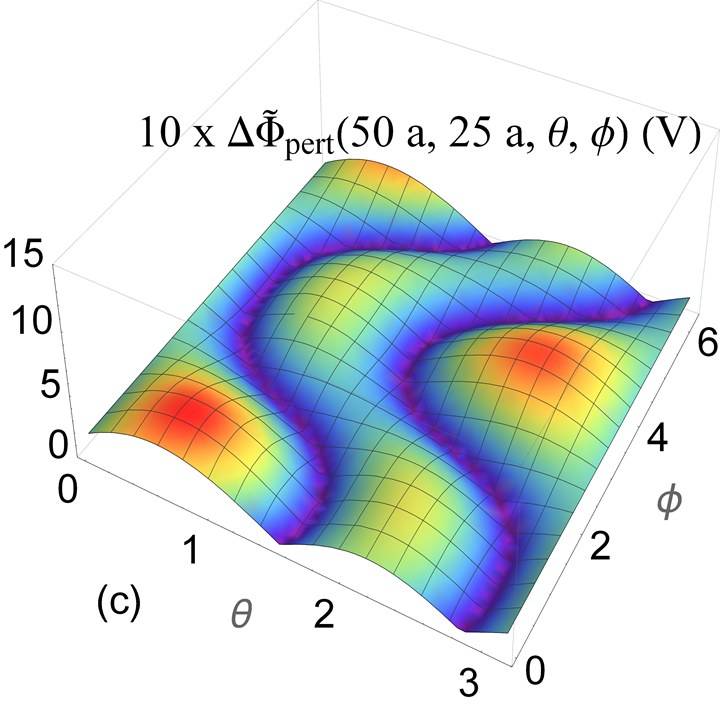} 
\includegraphics[width=0.2\linewidth]{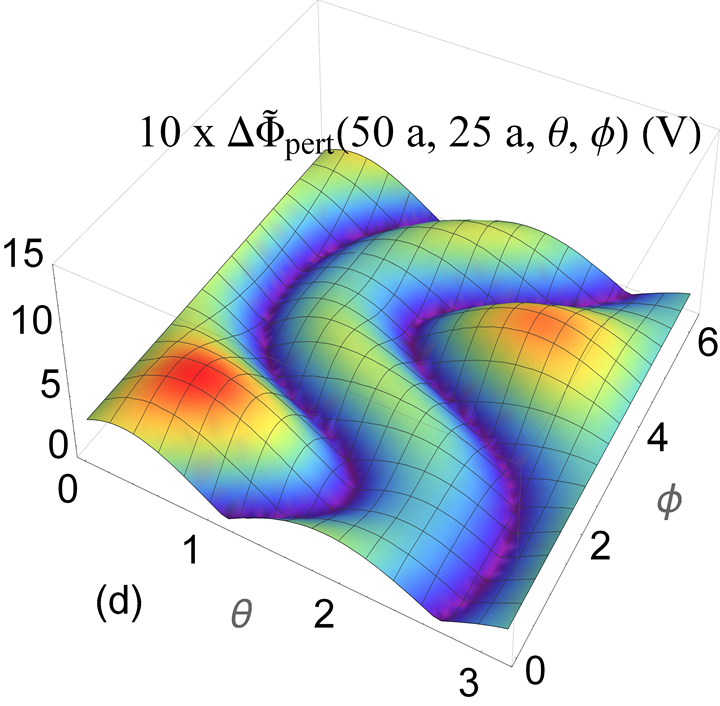}
\caption{As in Fig. \ref{c10}, but for $\Delta\tilde{\Phi}_\text{pert}(50a,25a,\theta,\phi)$.}
\label{c11}
\end{figure}

\begin{figure}[H]
 \centering
\includegraphics[width=0.2\linewidth]{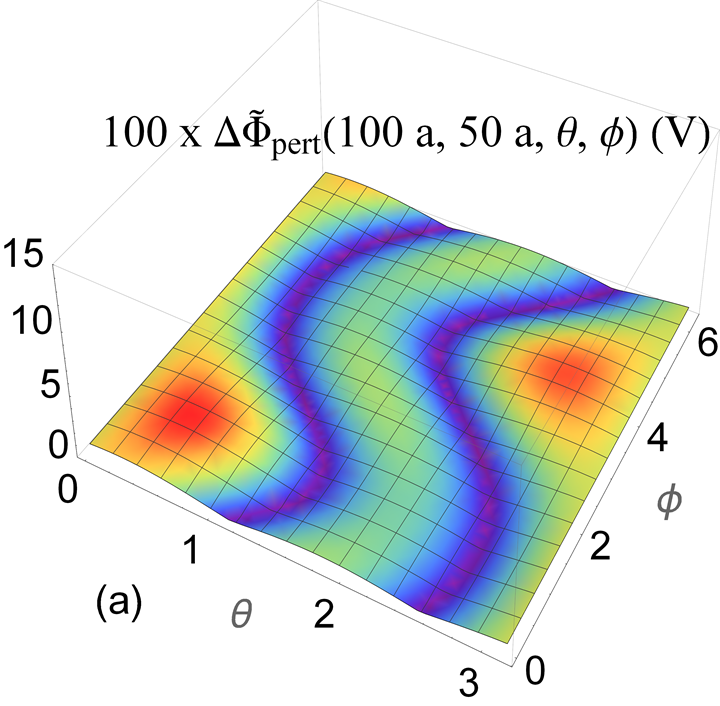} 
\includegraphics[width=0.2\linewidth]{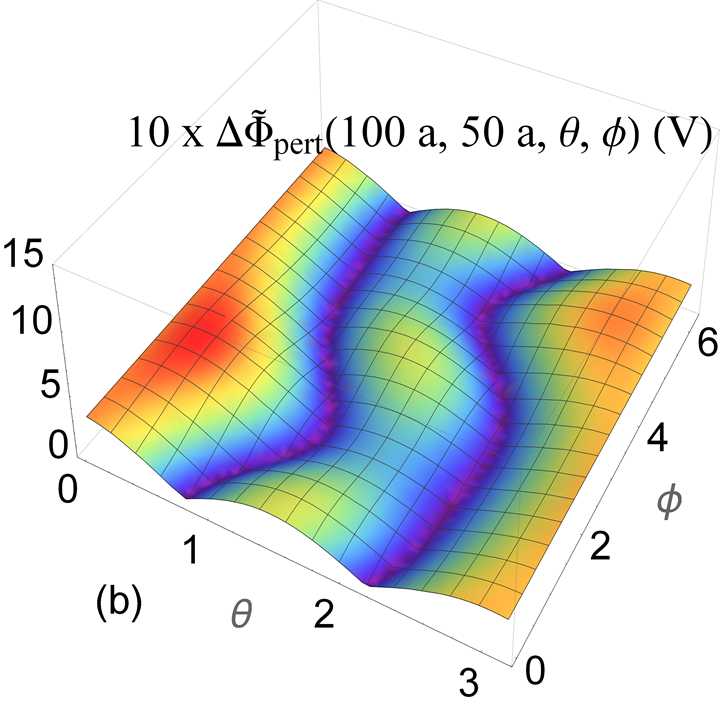} \\
\includegraphics[width=0.2\linewidth]{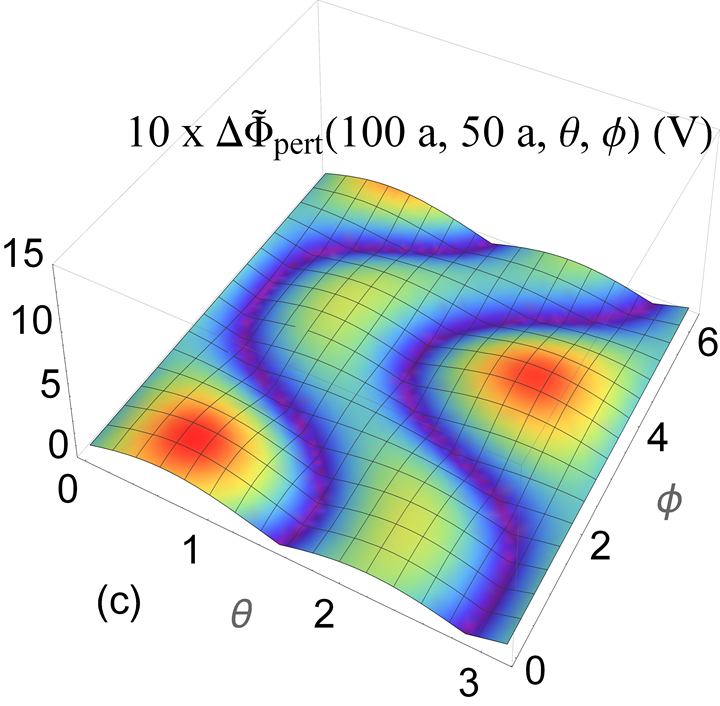} 
\includegraphics[width=0.2\linewidth]{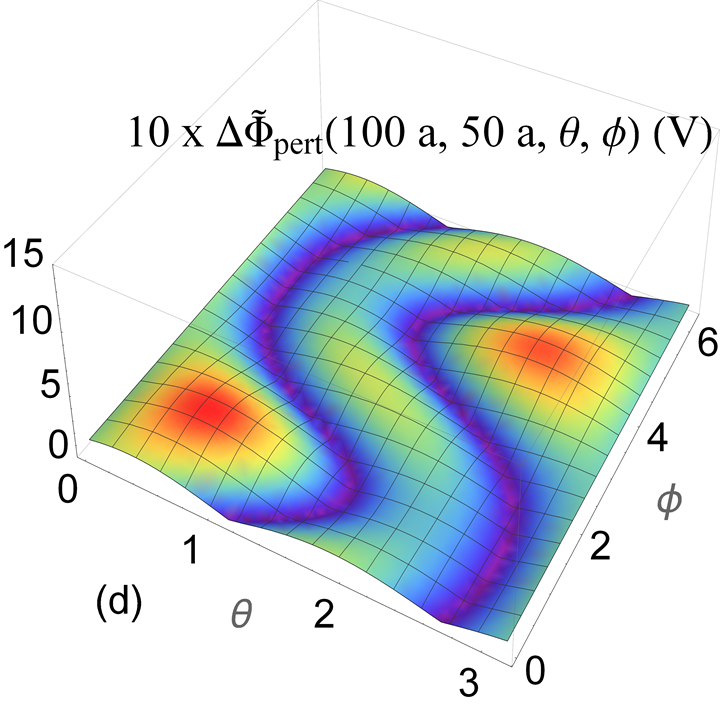}
\caption{As in Fig. \ref{c10}, but for $\Delta\tilde{\Phi}_\text{pert}(100a,50a,\theta,\phi)$.}
\label{c12}
\end{figure}

Unlike $\Fsca(\theta, \phi)$,  the normalized perturbation potential $\tilde{\Phi}_\text{pert}(r, \theta, \phi)$ depends  additionally on the distance $r>a$ from the origin to the observation point. With the same values of $a$, $\ro$, {$\thetao$,  $\phio$, $Q$, and $p$} as in Sec.~\ref{nrd2},
we also examined the perturbation potential's variations with respect to $\rtilde=r/a$, with $\alphax=1.2$ and $\alphay=1.6$ fixed, for a point-charge  source as well as for a point-dipole source with $\up \in\lec \ux,  \uy, \uz\ric$.
Figure \ref{c10} presents the angular profiles of {$\Delta\tilde{\Phi}_\text{pert}(25a,5a,\theta,\phi)$, 
Fig. \ref{c11} of   $\Delta\tilde{\Phi}_\text{pert}(50a,25a,\theta,\phi)$,   and Fig. \ref{c12} of 
 $\Delta\tilde{\Phi}_\text{pert}(100a,50a,\theta,\phi)$, where
 \begin{equation}
 \Delta\tilde{\Phi}_\text{pert}(r_b,r_a,\theta,\phi)= \vert
\tilde{\Phi}_\text{pert}(r_b,\theta,\phi)-\tilde{\Phi}_\text{pert}(r_a,\theta,\phi)\vert\,.
\end{equation}}
 
For all four sources, the inequalities 
{$\Delta\tilde{\Phi}_\text{pert}(25a,5a,\theta,\phi)
> \Delta\tilde{\Phi}_\text{pert}(50a,25a,\theta,\phi)> 
\Delta\tilde{\Phi}_\text{pert}(100a,50a,\theta,\phi)$ hold true. 
This indicates that, as $\rtilde$ increases, $\tilde{\Phi}_\text{pert}(r,\theta,\phi)$ decreases  as expected in every direction
 (indicated by $\lec\theta,\phi\ric$). Also, the surface of $\Delta\tilde{\Phi}_\text{pert}(100a,50a,\theta,\phi)$ is smoother and less undulating than the surface of $\Delta\tilde{\Phi}_\text{pert}(50a,25a,\theta,\phi)$, which is smoother and flatter than the surface of 
 $\Delta\tilde{\Phi}_\text{pert}(25a,5a,\theta,\phi)$.} This is due to the waning of the higher-order terms on the right side of Eq.~(\ref{Phi-sca-1}) as $\rtilde$ increases. These higher-order terms
 have a strong effect on $\Phi_\text{pert}(r,\theta,\phi)$ at observation points close to the sphere, the strongest such effects being observed for the dipole sources, as is evident from   Figs.~\ref{c10}(b)--(d). Indeed,  the decrease of $\tilde{\Phi}_\text{pert}(r,\theta,\phi)$ and the
 smoothening of its angular profile with increase of $\rtilde$ is reflected in the limit
\begin{equation}
\lim_{r\rightarrow \infty}\tilde{\Phi}\sca(r, \theta, \phi)=
\pi \,  \Phi\inc^{\rm ref} \,\Fsca(\theta, \phi)\,.
\end{equation}

Finally, on comparing Figs. \ref{c10}--\ref{c12}, we observe that the portions of the $\theta$-$\phi$ plane corresponding to the maximum/minimum increase in $|\tilde{\Phi}_\text{pert}(r,\theta,\phi)|$ remain almost the same as $\rtilde$ increases, for each of the four source potentials considered.

\section{Concluding Remarks}\label{s3}

We formulated and solved the boundary-value problem for the perturbation of an electric potential by a homogeneous  {anisotropic}  dielectric sphere in vacuum. As is commonplace for the exterior region, the source potential and the perturbation potential were represented
in terms of the standard solutions of the Laplace equation in the spherical coordinate system. A bijective spatial transformation
was implemented for the interior region in order to formulate the series representation of the internal potential.
Boundary conditions on the spherical surface were enforced and then the orthogonality of tesseral harmonics was employed
to derive
a transition matrix that relates
the expansion coefficients of the perturbation potential in the exterior region to those of the source potential.
The angular profile of the perturbation profile changes with distance from the center of the sphere, but eventually it settles down with
the   perturbation potential decaying as the inverse of the distance squared from the center of the sphere.

\vspace{5mm}
\noindent\textbf{Acknowledgement.} AL is grateful to the Charles Godfrey Binder Endowment at Penn State for ongoing support of
his research activities.

\vspace{5mm}

\end{document}